\documentclass[twocolumn,showpacs,preprintnumbers,amsmath,amssymb]{revtex4}

\usepackage{graphicx}
\usepackage{dcolumn}
\usepackage{bm}

\def\bea{\begin{eqnarray}}
\def\eea{\end{eqnarray}}

\begin{document}

\preprint{Version 2.6}

\title{Extrapolating parton fragmentation to low $Q^2$ in $e^+$-$e^-$ collisions}

\author{Thomas A. Trainor and David T. Kettler}
\address{CENPA 354290, University of Washington, Seattle, WA  98195}


\date{\today}

\begin{abstract}
We analyze the energy scale dependence of fragmentation functions from $e^+$-$e^-$ collisions using conventional momentum measures $x_p$ and $\xi_p$ and rapidity $y$. We find that replotting fragmentation functions on a normalized rapidity variable results in a compact form precisely represented by the beta distribution, its two parameters varying slowly and simply with parton energy scale $Q$. The resulting parameterization enables extrapolation of fragmentation functions to low $Q$ in order to describe fragment distributions at low transverse momentum $p_t$ in heavy ion collisions at RHIC.
\end{abstract}

\pacs{13.66.Bc, 13.87.-a, 13.87.Fh, 12.38.Qk, 25.40.Ep, 25.75.-q, 25.75.Gz}

\maketitle

\section{Introduction}

QCD theory predicts that an abundance of soft gluons (minijets) should be produced in relativistic collisions at RHIC~\cite{theor0}. Copious gluon production may drive formation of the colored medium in heavy ion collisions and global hydrodynamic phenomena~\cite{theor1,theor2}. However, the degree of equilibration of minijets in heavy ion collisions remains uncertain theoretically and experimentally. We should therefore search for and study remnants of {\em low}-$Q^2$ (energy scale $Q \sim$ 1 - 5 GeV) partons in single-particle and two-particle distributions of final-state hadrons. 


Measurements of {two-particle correlations} at RHIC with novel techniques have revealed substantial unequilibrated low-$Q^2$ parton fragment structure in p-p and Au-Au collisions~\cite{tomismd,axialci,mtxmt,mitpapers,hijsca,ptsca,lowq2,ppmeas,inverse,jeffqm04,jeffismd}. Fragment correlations have been measured in p-p collisions on transverse rapidity $y_t$ (defined below) and complementary angular subspace $(\eta,\phi)$ with {\em no jet hypothesis} (no high-$p_t$ trigger particle), providing access to fragments from {minimum-bias} partons (no analysis constraint on parton momentum) dominated by minijets. Jet correlations have been observed in p-p collisions for hadron $p_t$ down to 0.35 GeV/c (parton $Q \sim 1 - 2$ GeV)~\cite{jeffismd,tomismd}. Similar measurements in heavy ion collisions have revealed unexpected complexity. 


Initial measurements of two-particle angular correlations in p-p collisions at Fermilab (fixed target) and the CERN ISR on momentum subspace $(\eta,\phi)$ (pseudorapidity and azimuth)~\cite{etaphicorr} were described in terms of longitudinal (string) fragmentation~\cite{lund}. Jets---correlated fragments from hard-scattered partons---were first observed at larger $p_t$ and $\sqrt{s}$, establishing the nature of hard parton scattering and the validity of perturbative QCD (pQCD)~\cite{jets}.
A pioneering study of two-particle fragmentation functions in LEP $e^+$-$e^-$ collisions is described in~\cite{opalytxyt} ({\em cf.} a related theoretical treatment in~\cite{one-two-particle}). 


Minijet correlations in p-p and heavy ion collisions observed at RHIC represent QCD in a {\em non}-perturbative context: parton scattering, energy loss and fragmentation at low $Q^2$. We wish to connect those measurements to QCD theory {\em via} parton fragmentation measurements at larger $Q^2$ in elementary collisions. A context for {two}-particle fragment distributions at RHIC can be established by studying single-particle fragmentation functions (FFs) from p-\=p and $e^+$-$e^-$ collisions, the latter providing especially precise access to the fragmentation process down to very low parton $Q$ and hadron momentum. 


Low-$Q^2$ fragmentation is related to {\em local parton-hadron duality} (LPHD) which provides a correspondence between pQCD parton predictions and hadronic observables~\cite{dokshitzer}. According to LPHD conversion of partons to hadrons occurs locally in configuration space, with almost no distortion of the parton momentum distribution. The {structural} difference between a parton and a hadron should vanish for \mbox{$Q \sim$ 1 - 2~GeV} where parton production is {most abundant} in RHIC collisions. LPHD is important for low-$Q^2$ partons where `fragmentation' may terminate with one or two partons (and hence hadrons). Minijet-related minimum-bias two-particle correlations studied in p-p and A-A collisions~\cite{lowq2,axialci,tomismd,ppmeas,ptsca,jeffismd} may therefore provide details of parton fragmentation at the energy scale where LPHD is most important.


{\em Modification} of parton scattering and fragmentation in the QCD medium of heavy ion collisions may reveal properties of the medium itself. A theoretical study of in-medium modification of the single-particle FF in A-A collisions is reported in~\cite{borg}. The expectation is deformation of the {\em in vacuo} FF toward lower momentum, possibly corresponding to observed changes in the single-particle $p_t$ spectrum ratio $R_{AA}$ with collision centrality~\cite{raa,raa2}. A related study of two-particle correlations in heavy ion collisions, especially the asymptotic approach of fragment distributions to thermal equilibrium with increasing parton dissipation in the medium, is reported in~\cite{mtxmt}. Given the close connection between single-particle FFs and minijet-related two-particle fragment correlations, correlation measurements in heavy ion collisions may provide a more differential picture of properties of the QCD medium and its influence on low-$Q^2$ parton fragmentation.





In this paper we establish a basis for {\em extrapolation} of measured $e^+$-$e^-$ FFs to small energy scales as preparation for similar extrapolations in nuclear collisions. This is not a theoretical analysis based on pQCD methods. For a recent example of such an analysis which consistently describes FFs over a large $x_p$ (momentum fraction) range {\em cf.}~\cite{albino} and the related discussion in Sec.~\ref{kkpsec}. The present paper describes a phenomenological analysis of FF data intended to provide the best possible extrapolation down to small parton energies where pQCD assumptions such as collinearity and factorization become invalid.

The paper is organized as follows: We first present a new method of analyzing fragmentation functions, with emphasis on rapidity $y$ as a preferred kinematic variable for low-$Q^2$ fragmentation studies. We then consider the general properties of FFs in  the context of the double-log approximation (DLA), angular ordering and color coherence. We compare measured FFs from $e^+$-$e^-$ collisions at three energies on several momentum variables and describe a new form of approximate energy-scale invariance on {\em normalized rapidity} $u$. We demonstrate that FFs on $u$ are precisely modeled by the {\em beta distribution}. We consider FFs for identified hadrons and identified partons. Based on fits to measured FFs and jet multiplicity data we develop a simple, precise parameterization of $e^+$-$e^-$ FFs valid over a broad energy range. Finally, we use our parameterization to study scaling violations and extrapolation to low $Q^2$.

\section{Analysis Method} \label{sec3}

The FF $D(x_E,Q^2)$ as used in this analysis is a single-particle density $2dn/dx_E$ of hadron fragments on energy fraction $x_E = E_{hadron} / E_{parton}$ produced by a pair of partons (dijet) with total energy $Q$ ($Q^2 = -q^2$ is the negative invariant mass squared for the initial momentum transfer). Momentum fraction $x_p = p_{hadron}/p_{parton}$ approximates $x_E$ if particle momenta are measured. At large $x$ the distribution shape reflects energy conservation during the parton splitting cascade. At small $x$ the shape is determined by quantum coherence of gluon emission (gluon or color coherence and the hadron size scale)~\cite{frag,coher}. The FF data used in this study are hadron distributions reported on momentum fraction $x_p, $ or $\xi_p \equiv \ln(1/x_p)$. Distributions on $x_p$ emphasize pQCD aspects of parton fragmentation at large $p$ ({\em e.g.,} scaling violations). For non-pQCD effects, especially the role of gluon coherence, logarithmic variable $\xi_p$ provides better visual access to the relevant small-$x_p$ (large-$\xi_p$) region. 

$D_p^h(x,Q^2 \text{ or } s)$ is the FF for parton type $p$ and hadron type $h$ at the energy scale denoted by $Q^2$ or $s$. The parton-flavor-inclusive distribution $D^h(x,s)$ is discussed in Sec.~\ref{hadron-id}, and the fragment-flavor-inclusive distribution $D_p(x,s)$ is discussed in Sec.~\ref{parton-id}. The total FF is $D(x,s) = \sum_h D^h(x,s)$. The corresponding FF on $\xi$ is $D(\xi,s) = x\, D(x,s)$, with Jacobian factor $x$. FFs satisfy relations $\int_0^1 dx\, D(x,s) =  2n$ ( dijet fragment multiplicity) and $ \int_0^1 dx\, x\, D(x,s) = 2$ (dijet energy conservation)~\cite{biebel}. To simplify notation we adopt the convention that symbol $D$ represents any fragmentation function, with the specific form [Jacobian relation to $D(x,s)$] implied by the first argument. Plot axes are labeled with the corresponding dijet particle density $D(x,s) \rightarrow 2dn/dx$, $D(\xi,s) \rightarrow 2dn/d\xi$, {\em etc.} to avoid confusion.


This study focuses on low-$Q^2$ parton fragmentation. Since $\xi_p$ and pQCD expansion parameter $Y(Q) = \ln(Q/\Lambda)$ ($\Lambda$ represents a reference energy scale) are undefined as $p$, $Q \rightarrow 0$ we introduce rapidity $y$ (well-behaved in that limit) as an alternative logarithmic momentum/energy variable. The rapidity along axis $\hat z$ is $y_z(\vec{p};m_0) \equiv \ln[(E + p_z)/m_t]$, with transverse (to $\hat z$) mass $m_t^2 = m_0^2 + p_t^2$. In frames comoving on $\hat z$ $\vec{p} \rightarrow p_t$, $E \rightarrow m_t$ and $y \rightarrow y_t = \ln\{(m_t + p_t)/m_0\}$. In a frame where $p$ is the only non-zero momentum component $y(p;m_0) = \ln[(E + p)/m_0]$, with $y \rightarrow \ln(2p/m_0)$ for $p \gg m_0$ and $\rightarrow p/m_0$ for $p \ll m_0$. $m_0$ may be a quark or hadron mass or energy scale $\Lambda$. 


Given the limiting cases for $y$ we note that $\ln(\sqrt{s}/m_0) \sim y(\sqrt{s}/2;m_0) \equiv y_{max}$, the {kinematic limit} for fragment rapidities. Similarly, $Y(Q) = \ln(Q/\Lambda) \sim y(Q/2;\Lambda)$ is a rapidity measure of the energy scale relative to a reference scale. We observe for data a lower limit $y_{min}$ which may depend on fragment species and collision system ($e^+$-$e^-$ {\em vs} p-\=p). For unidentified fragments we assign the pion mass $m_0 \rightarrow m_\pi$ to all hadrons (but {\em cf.} Sec.~\ref{hadron-id}). 
From data distributions on $x_p$ or $\xi_p$ for parton energy scale $Q$ or CM energy $\sqrt{s}$ we extract fragment momenta $p$ and calculate equivalent rapidities $y$ (fragments) and $y_{max}$ (partons). Data distributions on $x_p$ or $\xi_p$ are transformed to distributions on $y$ using appropriate Jacobians. In~\cite{alphs-delph} $\ln(1/x_p) \rightarrow \ln\{(E + p)_{parton} / (E + p)_{hadron}\} = y_{max} - y$ exactly.





\label{beta0}
 
Most $e^+$-$e^-$ FFs plotted on normalized rapidity $u \equiv (y - y_{min}) / (y_{max} - y_{min}) \approx 1 - \xi_p / Y$ have a particularly simple form described by the {\em beta distribution}. The unit-normal beta distribution defined on $u \in [0,1]$ is $\beta(u;p,q) = u^{p-1}\, (1-u)^{q-1} / B(p,q)$, with parameters $p,\,q \geq 0$ and beta function $B(p,q) = \frac{\Gamma(p)\, \Gamma(q)}{\Gamma(p+q)}$. Parameters $p$ and $q$ determine the shape of the distribution below and above the mode (most probable point) respectively. The mode is $u^* = \frac{p-1}{p+q-2}$, the mean is $\bar u = \frac{p}{p+q}$ and the variance is $\sigma_u^2 = \frac{p\, q}{(p+q)^2(p+q+1)} \approx \frac{1}{4(p+q+1)}$ (to 2\%).


\section{$e^+$-$e^-$ Fragmentation Functions}


 
\begin{figure}[h]
\includegraphics[width=3.3in,height=1.68in]{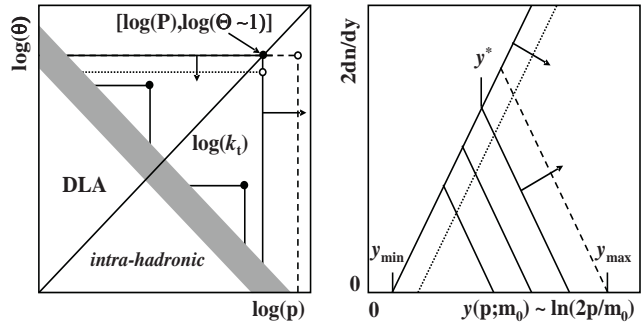}
\caption{\label{fig1}
Left panel: Schematic illustration of the double-log approximation (DLA) to parton fragmentation. Right panel: Self-similar variation with energy of the the corresponding fragment distribution on $y$. 
}
\end{figure}

The double log approximation (DLA~\cite{coher}) 
provides a context for extrapolating the fragmentation process to low $Q^2$. The fragment emission probability is approximated by a uniform density on logarithmic space $[\log(\theta),\log(p)]$, where $\theta$ is the radiated-parton emission angle and $p$ is its momentum. The distribution is sketched in Fig.~\ref{fig1} (left panel), where $P$ 
is the leading-parton momentum and $\Theta \sim 1$ is the jet angular acceptance. The large solid triangle represents the kinematic boundary for the first radiated parton. The smaller triangles illustrate the {\em self-similar} nature of the splitting process (angular ordering~\cite{angorder}), each radiated parton becoming itself a DLA radiator. Alternatively, the DLA may be expressed in terms of $d\log(k_t)\, dy$, where $k_t$ is the transverse momentum component relative to the radiating parton momentum, and $y$ is the radiated parton rapidity~\cite{lund-dla}. The flat DLA emission probability is terminated at some $k_t$ (grey band in Fig.~\ref{fig1} left panel) due to {\em gluon coherence}~\cite{coher}. For sufficiently small $k_t$ the conjugate transverse size of the virtual gluon overlaps the radiating parent parton and parton showering is terminated. 


The general form of the FF and its evolution with $p_{parton}$ corresponding to the DLA with angular ordering and gluon coherence (modified leading log approximation or MLLA~\cite{mlla}) is sketched in Fig.~\ref{fig1} (right panel). 
We expect a monotonic increase with decreasing $y$ below $y_{max}$ due to showering. The available phase space (above the band in the left panel) is however reduced with decreasing $y$ by gluon coherence, causing the FF to turn over, with a maximum at $y^*$ (the mode). The distribution then falls to zero at some $y_{min}$ which may be nearly independent of $y_{max}$. 
The FF is apparently self-similar at two levels: the internal cascade and its external boundary. Reducing the maximum opening angle $\Theta$ (dash-dot line) or increasing the parton momentum $P$ (dashed line) changes the boundary of the cascade (left panel) and correspondingly the FF (right panel).


\section{Fragment Distribution on $x_p$ and $\xi_p$}

\begin{figure}[h]
\includegraphics[width=1.65in,height=1.79in]{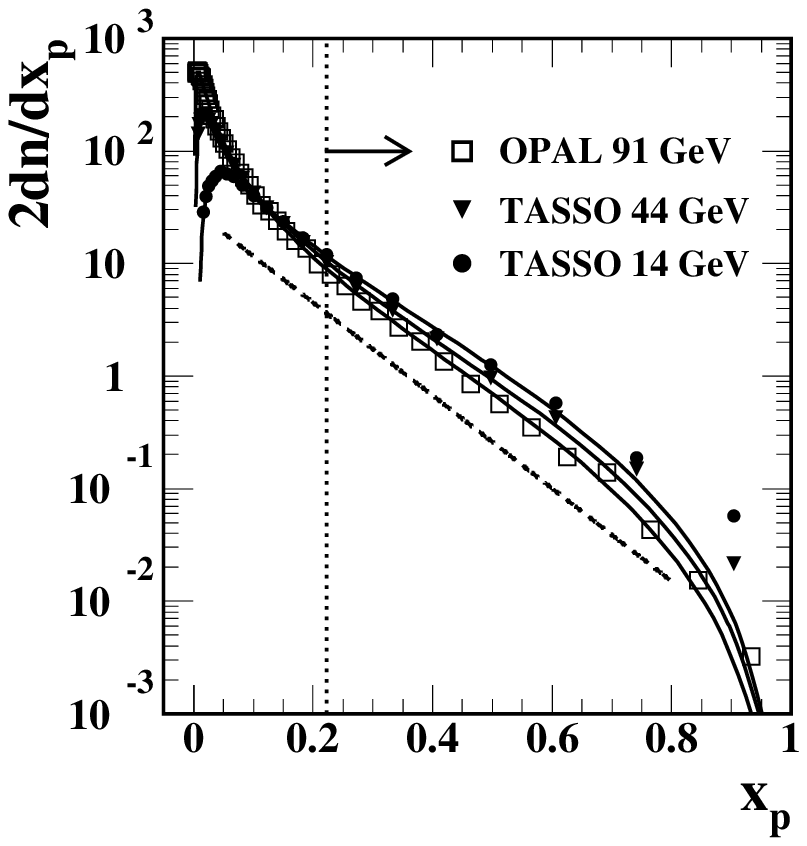}
\includegraphics[width=1.65in,height=1.75in]{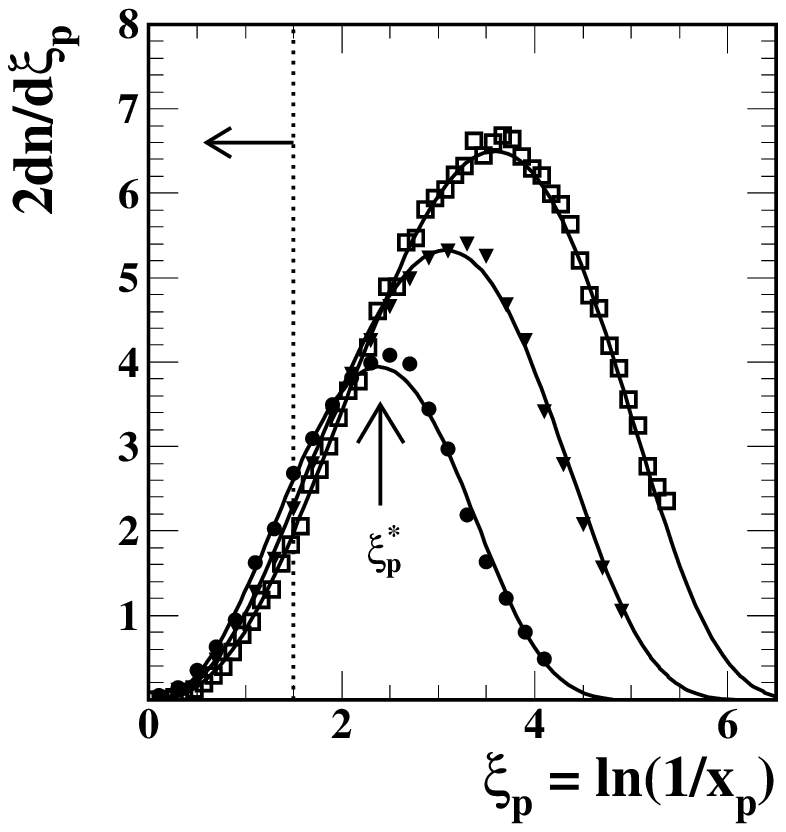}
\caption{\label{fig2}
Left panel: $e^+$-$e^-$ fragmentation functions on fractional momentum $x_p = p_\text{fragment} /p_\text{parton}$ for three CM energies. The dashed line is an exponential reference. Right panel: The same fragmentation functions on logarithmic variable $\xi_p = \ln (1/x_p)$. The vertical dotted lines mark equivalent points on the two variables. The solid curves are determined by the parameterization from this analysis. 
}
\end{figure}

Single-particle FFs from $e^+$-$e^-$ collisions have been studied extensively~({\em e.g.,} \cite{tasso,frag,cdf,cdf2,opal-boost,opal-sing,cdf-glue}). The FF data plotted in Figs.~\ref{fig2} - \ref{fig5} were obtained from collisions at three energy scales (CM energy $Q = \sqrt{s}$ = 14, 44 and 91.2 GeV) measured at PETRA~\cite{tasso} and LEP~\cite{frag} for unidentified hadrons from unidentified partons (flavor-inclusive jets). Those distributions are fiducial for this study because of the exceptional data quality and fragment momentum coverage. We consider the data in several presentation schemes and then develop a parameterized representation for extrapolation to low $Q^2$. 

In Fig.~\ref{fig2} we plot fragment distributions on momentum fraction $x_p$ (left panel) and logarithmic equivalent $\xi_p$ (right panel). Distributions on $x_p$ emphasize the large-$x_p$ (small-$\xi_p$) region where pQCD is expected to best describe data, where the na\"ive parton model predicts `scaling' or invariance of the parton distribution on energy scale $Q$. The dashed reference line in the left panel illustrates the exponential model sometimes used to characterize FFs on $x_p$. The vertical dotted line corresponds to $\xi_p = 1.5$ in the right panel: only a small fraction of fragments fall above that point. The data exhibit systematic {scaling violations} ($Q$ dependence) described by the DGLAP evolution equations ({\em cf.} Sec.~\ref{scalviol})~\cite{dglap,dglap2}. To study scaling violations FFs on $x_p$ are parameterized by a model function such as $D_{p}^h(x,Q^2) = N\, x^\alpha\, (1-x)^\beta\, (1+\gamma / x)$, where the four parameters depend on parton type $p$, hadron type $h$ and energy scale $Q$~\cite{webber-xp-param,kkp}. Distribution details in the small-$x_p$ region ({\em e.g.,} below $x_p = 0.1$) are minimized in this format ({\em cf.} Sec.~\ref{kkpsec}). 

We can also plot data on $\xi_p$ (right panel) which emphasizes the small-$x_p$ (large-$\xi_p$) region and better reveals non-perturbative details of fragmentation. The distribution is {\em approximately} gaussian, with mode $\xi^*_p$ and r.m.s.~width $\sigma_{\xi_p}$ predicted by pQCD ({\em cf.} Sec.~\ref{ximode}). As noted, the fall-off at large $\xi_p$ and maximum at $\xi_p^*$ result from {gluon coherence}~\cite{frag,coher}. Measurement of the {full} fragment distribution above and below the mode is important for a complete characterization of the fragmentation process.


The solid curves in Figs.~\ref{fig2} - \ref{fig5} are obtained from beta distributions on normalized rapidity $u$ determined by the systematic trends of beta parameters $(p,q)$ plotted in Fig.~\ref{fig8} (left panel) ({\em cf.} Sec.~\ref{betasum}) and transformed to each plotting space with appropriate Jacobians. Some approximation to `scaling' or energy-scale independence is expected at large $x_p$ ({small} $\xi_p$). Another form of scaling at small $x_p$ ({large} $\xi_p$) may be explored by plotting distributions on rapidity $y$.

\section{Fragment Distribution on $y$}

\begin{figure}[h]
\includegraphics[width=1.65in,height=1.75in]{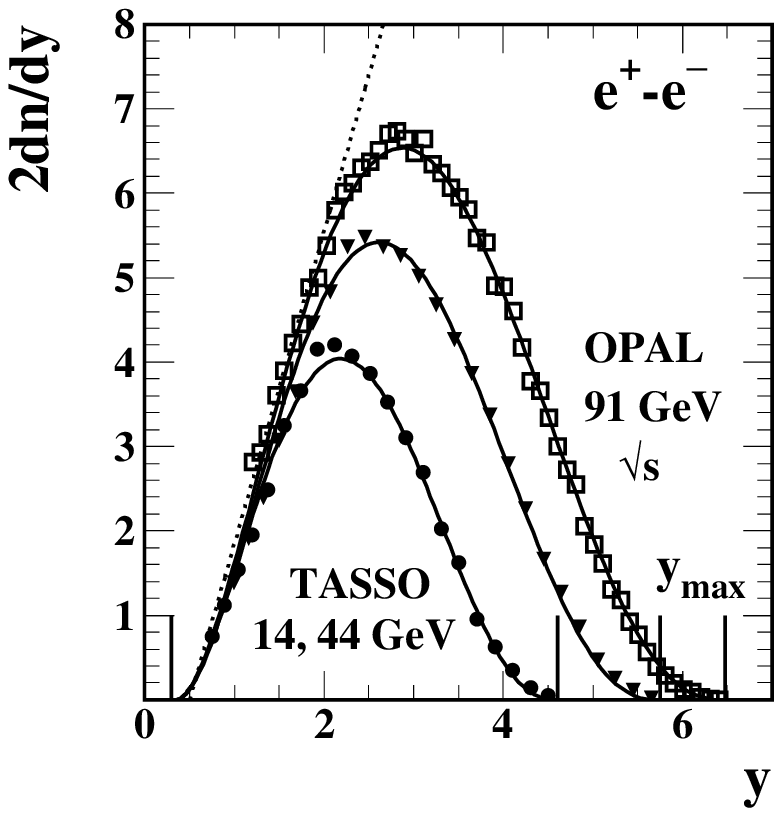}
\includegraphics[width=1.65in,height=1.75in]{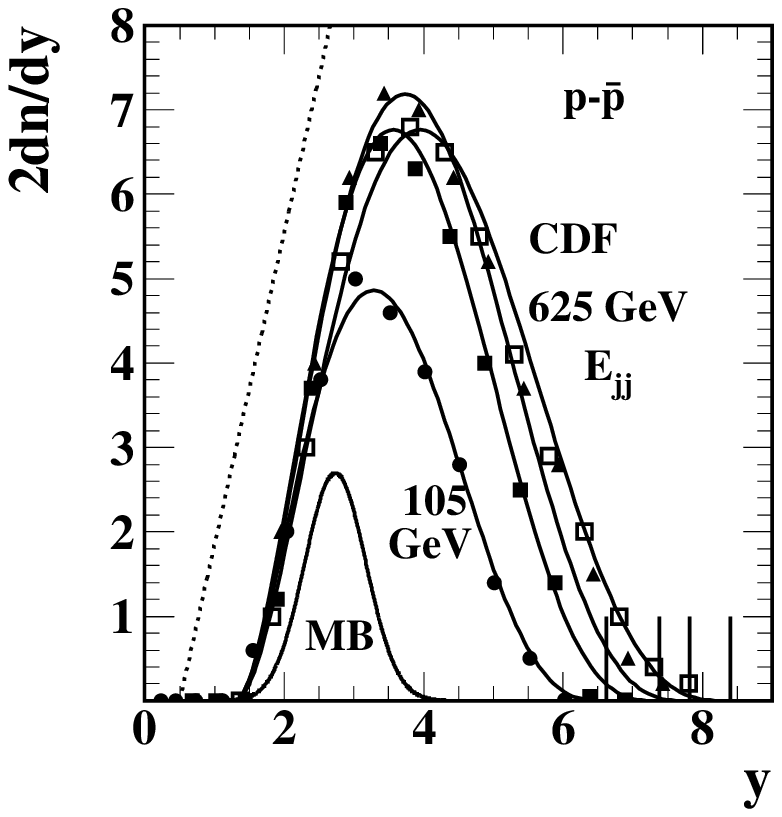}
\caption{\label{fig3} Fragmentation functions on rapidity $y$ for $e^+$-$e^-$ collisions (left panel) and p-\= p collisions (right panel). 
}
\end{figure}

Fragmentation functions plotted on $\xi_p$ coincide at the kinematic limit $\xi_p = 0$  corresponding to the parton momentum. 
However, in Fig.~\ref{fig3} (left panel) we observe that the FFs for three energies plotted on $y$ have a common low-momentum limit $y_{min} \sim 0.35$ (vertical line, and {\em cf.} Sec.~\ref{ymin}). That alignment is possible because $y$ has the well-defined limiting value 0 as momentum $p \rightarrow 0$. 
Each data FF is terminated at the upper end by its kinematic limit $y_{max} = y(\sqrt{s}/2;m_0)$ (vertical lines) corresponding to $\xi_p = 0$ in Fig.~\ref{fig2}. The distribution maxima increase monotonically with collision energy. The FFs in the left panel illustrate the self-similarity sketched in Fig.~\ref{fig1} (right panel) and confirm an expectation for DLA scaling: fragmentation at small $y$ should be nearly independent of the leading parton momentum.

In Fig.~\ref{fig3} (right panel) we plot FFs from p-\=p collisions at FNAL~\cite{cdf,cdf2} (the points are {samples} from the original data distributions used here to illustrate qualitative features). While the general features are similar to FFs for $e^+$-$e^-$ collisions the lower limit $y_{min}$ is considerably larger for p-\=p collisions (note the dotted reference line common to the two panels). 
The larger $y_{min}$ for p-\=p collisions ($\sim 1.5$) may be due to the finite jet-cone opening angle~\cite{cdf2} and/or the presence of the underlying event~\cite{underly1} which must be distinguished from jet fragments. The gaussian curve labeled MB represents a {\em minimum-bias} fragment distribution (no selection is imposed on the parton momentum spectrum) derived from the event-multiplicity dependence of p-p $p_t$ spectra~\cite{pptwocomp} which compares well with the systematics of FFs obtained from p-\=p jet reconstruction. 

\begin{figure}[h]
\includegraphics[width=3.3in,height=1.75in]{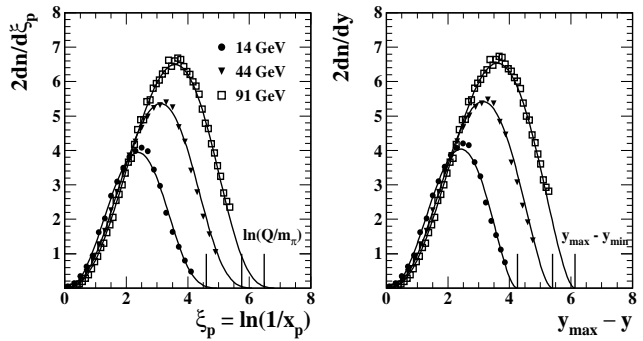}
\caption{\label{fig4}
Comparison of fragmentation functions for three energies on $\xi_p$ and $y_{max} - y$. Differences are noticible only for small $y$ or $p$ (large $\xi_p$).  
}
\end{figure}

In Fig.~\ref{fig4} we compare FFs on $\xi_p \equiv \ln(Q/2p)$ (left panel) and $y_{max} - y \sim \ln(Q/2p)$ (right panel). The distributions are equivalent below the upper half-maximum points ($p \gg m_0$), above which distributions on $y_{max} - y$ drop rapidly toward well-defined limits at $y_{max} - y_{min}$. Distributions on $\xi_p$ extend in principle to $\infty$, but the transformed beta distributions limit at $p_{min} \sim m_\pi / 2$ or $\xi_p \sim \ln(Q/m_\pi)$, indicated by vertical lines in Fig.~\ref{fig4} (left panel). This comparison suggests that rapidity $y(Q/2;m_0)$ or difference $y_{max} - y$ could replace $\xi_p$ in FF studies. Rapidity $y(Q/2;\Lambda)$ could also replace pQCD expansion parameter $Y(Q) = \ln(Q/\Lambda)$, remaining well-defined for $Q \rightarrow 0$ while preserving established pQCD relations for larger $Q$. Figs.~\ref{fig3} and \ref{fig4} also suggest that rescaling the rapidity by $y_{max} - y_{min}$ might provide more differential access to FFs.

\section{Fragment Distribution on $u$} \label{betau}

\begin{figure}[h]
\includegraphics[width=3.3in,height=1.75in]{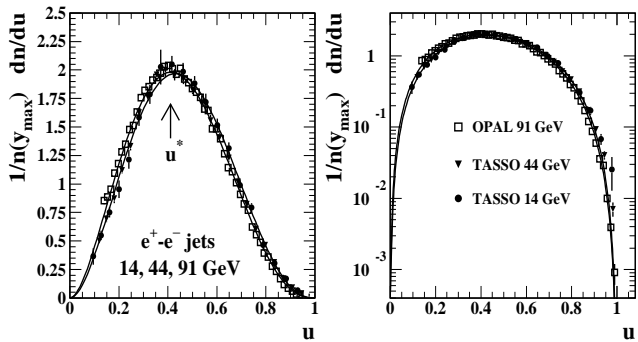}
\caption{\label{fig5}
Fragmentation functions plotted on normalized rapidity $u$ in linear (left panel) and semi-log (right panel) formats. The data distributions have been normalized by the corresponding di-jet multiplicity at each energy (lower solid curves in Fig.~\ref{fig7}) determined by parameters $(p,q)$. The data for three energies are plotted, but the curves for only 14 and 91 GeV are plotted to provide visible separation. 
}
\end{figure}

Expectations of approximate energy scaling at large $x_p$ and a different form of scaling (gluon coherence) at small $x_p$ seem to require conflicting plotting strategies on $\xi_p$ and $y$. However, both forms can be accommodated 
with normalized rapidity $u \equiv (y - y_{min}) / (y_{max} - y_{min}) \in [0,1]$. FFs from $e^+$-$e^-$ collisions can be factored as $D(u,y_{max}) = 2n(y_{max})\, g(u,y_{max})$, with dijet multiplicity $2n(y_{max})$ ({\em cf.} Fig.~\ref{fig7}) and unit-normal form factor $g(u,y_{max})$. In Fig.~\ref{fig5} we plot the three representative FFs transformed to $1/n(y_{max})\, dn/du \equiv g(u,y_{max})$. Multiplicity $2n(y_{max})$ can be obtained from fits to data, but also from the {\em shape} of $g(u,y_{max})$ ({\em cf.} Sec.~\ref{jetmult}).

We have determined that data form factor $g(u,y_{max})$ is well-described by beta distribution $\beta(u;p,q)$ defined in Sec.~\ref{beta0}. 
While there are substantial `scaling violations' on $x_p$ or $\xi_p$ ({\em cf.} Sec.~\ref{scalviol}), the normalized FF shapes on $u$ in Fig.~\ref{fig5} are {nearly} independent of $Q^2$ or $y_{max}$ over a substantial energy range. However, the remaining small variations with energy are significant, and well described by energy-dependent beta parameters $(p,q)$ plotted in Fig.~\ref{fig8} and discussed in Sec.~\ref{betasum}. 





\section{Identified Hadron Fragments} \label{hadron-id}

\begin{figure}[h]
\includegraphics[width=1.65in,height=1.75in]{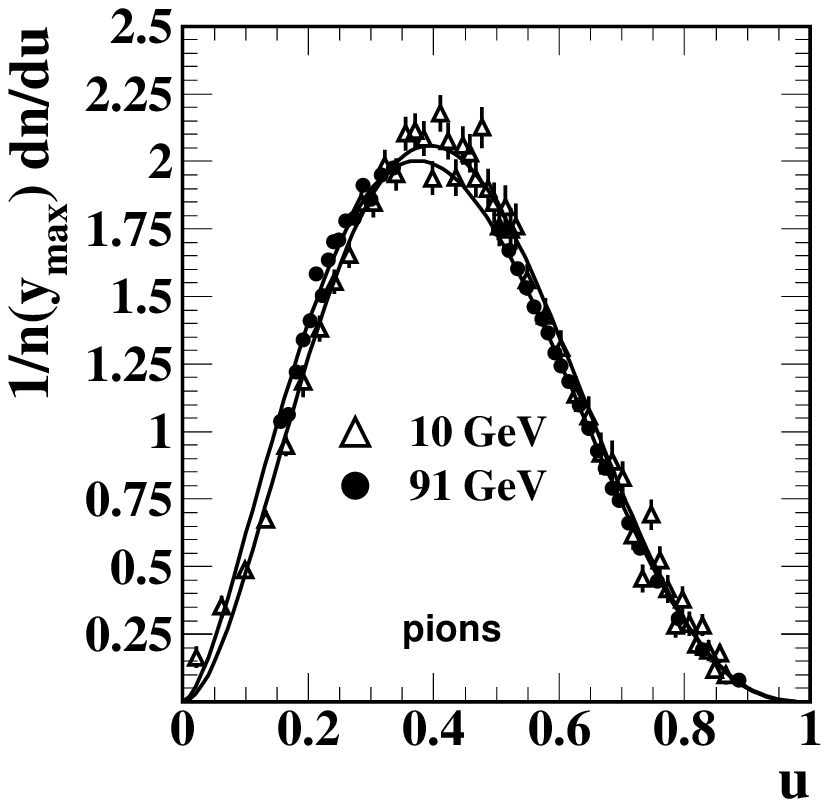}
\includegraphics[width=1.65in,height=1.75in]{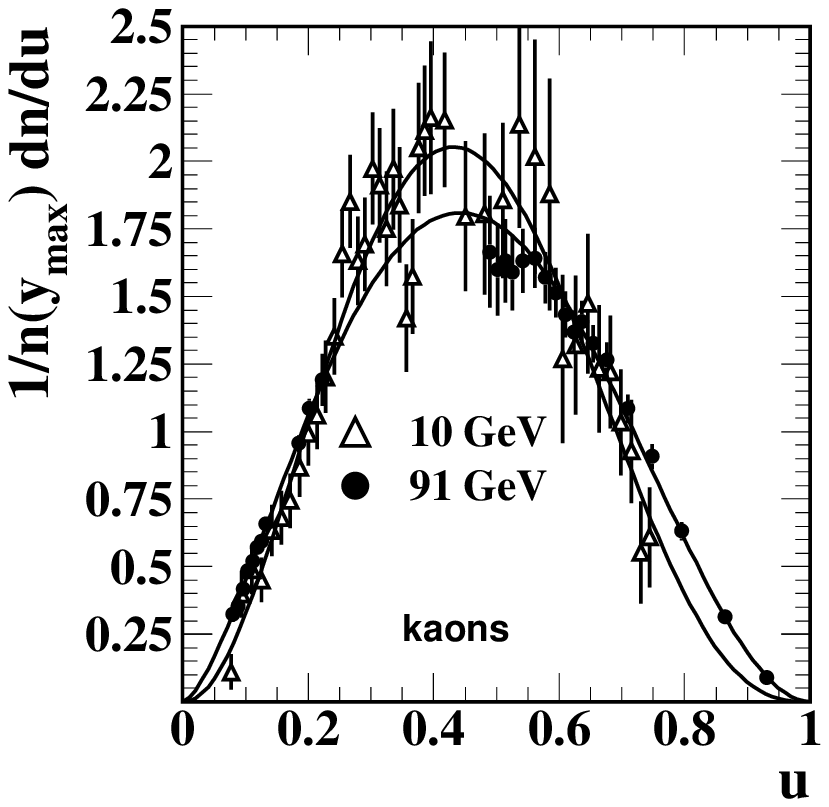}
\caption{\label{fig10}
Fragmentation functions for two CM energies and for pion (left panel) and kaon (right panel) fragments plotted on normalized rapidity $u$. 
}\end{figure}

We define rapidities for unidentified hadrons by assigning the pion mass $m_0 \rightarrow m_\pi$ to several particle species. To assess the consequences we use identified-particle FF data for two CM energies. In Fig.~\ref{fig10} we show data $g(u,y_{max})$ and best-fit model $\beta(u;p,q)$ for identified charged pions $\pi^\pm$ (left panel) and kaons K$^\pm$ (right panel) at 10 GeV~\cite{pid-10gev} and 91 GeV~\cite{pid-91gev}. Parton rapidity $y_{max}$ is determined in each case with the identified {\em hadron fragment mass}. The distributions for identified protons (p,\=p) show similar behavior but with larger statistical errors. The pion FFs have widths similar to unidentified hadrons, but the peak modes are significantly lower (0.38 {\em vs} 0.41 at 91 GeV). The kaon peak modes are comparable to those for unidentified hadrons but the peak width at higher energy is significantly larger. The kaon FF shape seems to converge on the pion distribution at lower energy. The apparent blending of quark flavors could be related to the convergence of the gluon and quark FFs at lower energy in Fig.~\ref{fig8}.

\begin{figure}[h]
\includegraphics[width=3.3in,height=1.75in]{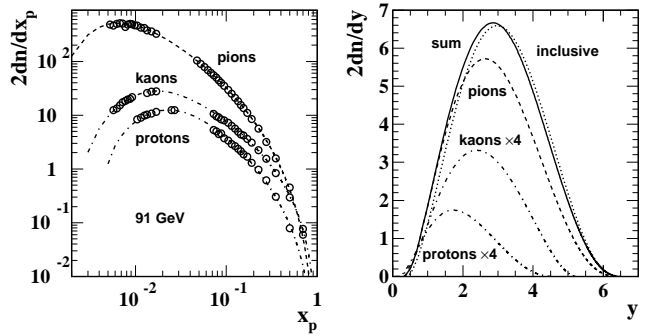}
\caption{\label{fig11}
Left panel: Fragmentation function data for $\sqrt{s}$ = 91 GeV and for three hadron species on fractional momentum $x_p$ and corresponding beta-distribution fits on normalized rapidity $u$ transformed to $x_p$ (dash-dot curves). Right panel: Dashed and dash-dot curves for identified hadrons with proper mass assignments transformed from the left panel. The solid curve is a sum on $x_p$ of the beta-distribution fits from identified fragments in the left panel transformed to rapidity $y$ with a pion mass assignment. The solid curve is close to the distribution obtained from inclusive hadrons (dotted curve). 
}
\end{figure}

To determine the effect of assigning the pion mass to unidentified fragments we used the following procedure. Data distributions on $x_p$ for three identified fragment species (the 91 GeV data in Fig.~\ref{fig10}) were transformed to normalized rapidity $u$ with the proper mass assignments.  Functions $\beta(u;p,q)$ were fitted to each species, transformed back to $x_p$ and plotted (dash-dot curves) with the data in Fig.~\ref{fig11} (left panel). The model functions on $x_p$ were summed to represent the combination of unidentified hadrons and transformed to rapidity $y$ assuming the pion mass, giving the solid curve in Fig.~\ref{fig11} (right panel). The dotted curve was obtained by assigning the pion mass to all data, transforming to $u$, fitting the resulting distribution and then transforming back to $y$. We conclude from the results that misidentifying kaons and protons as pions in unidentified hadrons shifts the FF peak mode at 91 GeV from the pion value $\sim$ 0.38 to the inclusive hadron value $\sim$ 0.41 in Fig.~\ref{fig5} (left panel). The dashed and dash-dot curves are fits to the individual fragment species with proper masses used to determine the rapidities. 

From this exercise certain trends are notable: Proton fragments have the largest momenta but the smallest rapidities. When transformed to normalized rapidity $u$ the FFs for different fragment species are similar in shape (beta distribution) but exhibit small but significant mode variations with parton energy and hadron species ({\em cf.} Fig.~\ref{fig13} and the discussion following for a summary of flavor dependence). Unit-normal data distributions $g(u,y_{max})$ for all light hadron species are well-described by model $\beta(u;p,q)$, establishing applicability of the beta distribution to FFs for identified light meson and baryon fragments as well as to inclusive hadrons.

\section{Identified Partons} \label{parton-id}

\begin{figure}[h]
\mbox{
\includegraphics[width=1.65in,height=1.65in]{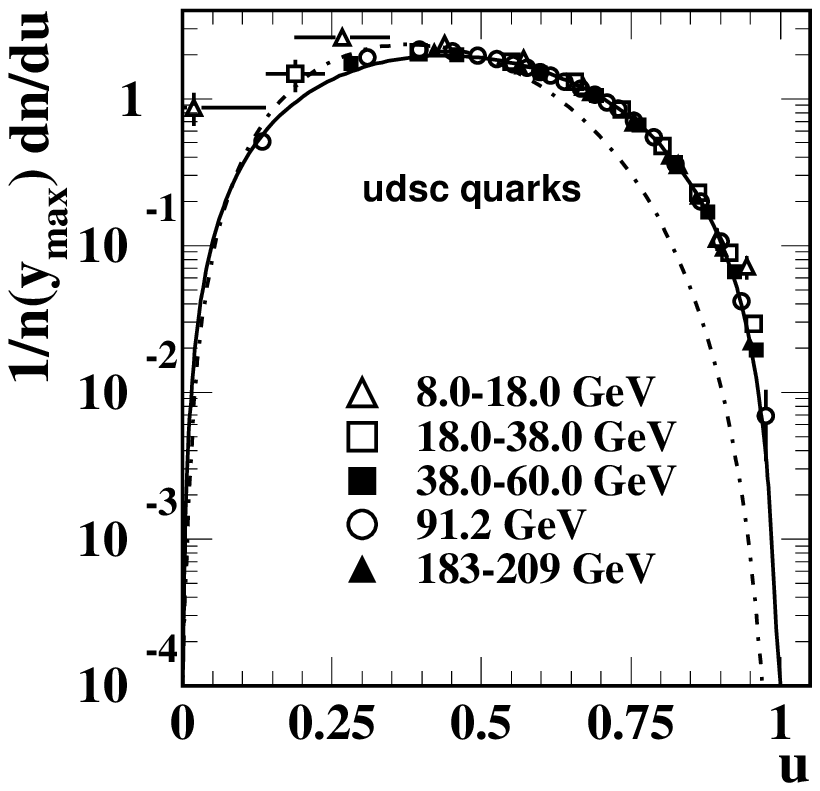}
\includegraphics[width=1.65in,height=1.65in]{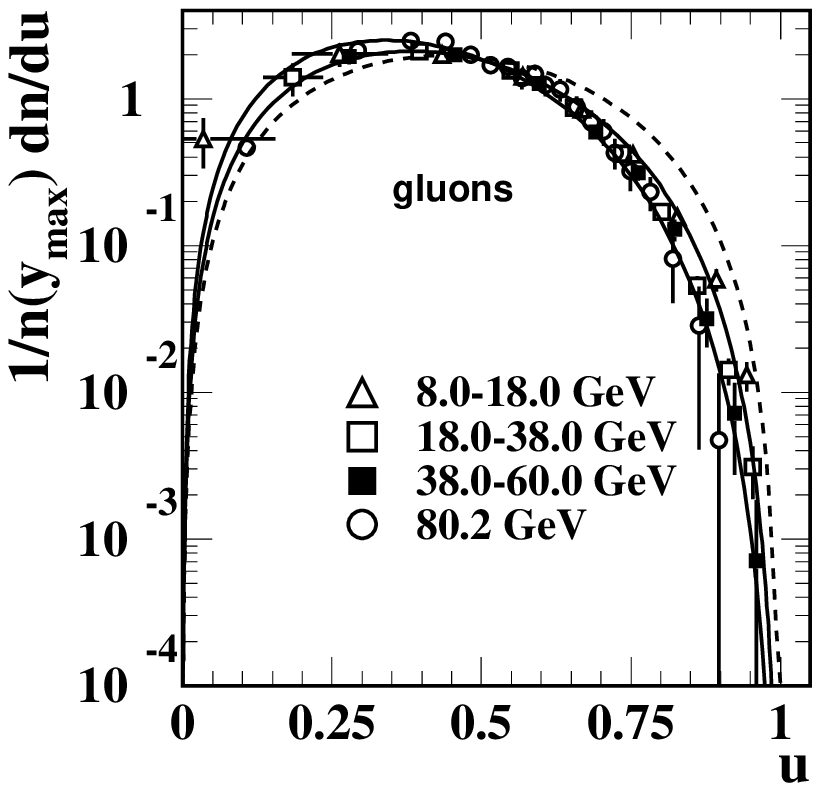}
}
\mbox{
\includegraphics[width=1.65in,height=1.65in]{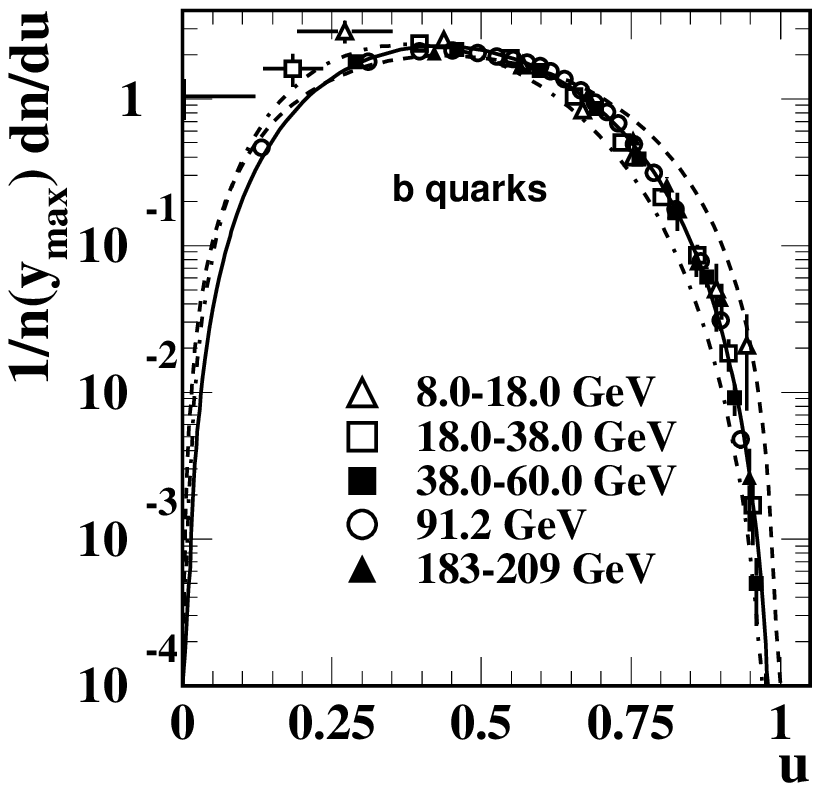}
\includegraphics[width=1.65in,height=1.68in]{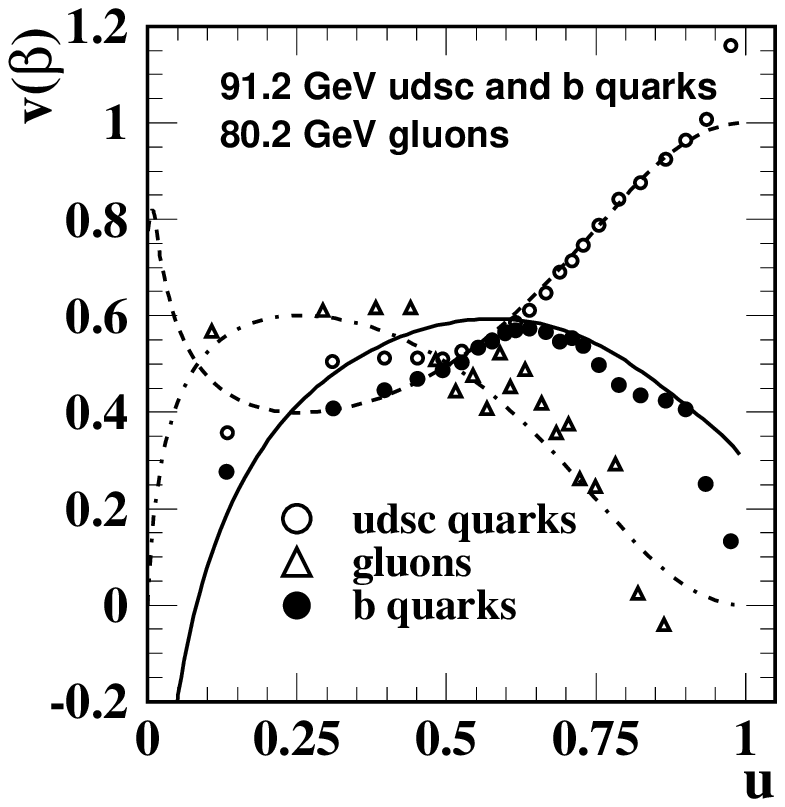}
}
\caption{\label{fig12}
Unit-normal fragment distributions on normalized rapidity $u$ for udsc quarks (upper left), b quarks (lower left) and gluons (upper right) fragmenting to inclusive hadrons. Lower-right panel: Normalized logarithmic variable $v(\beta)$ (see text) measures the FF shape for b-quark jets (solid) relative to those for gluon jets (dash-dot) and udsc-quark jets (dashed) as references.  
}
\end{figure}

We now consider the role of parton identity in FF systematics. Normalized data distributions on $u$ are shown in Fig.~\ref{fig12} for inclusive hadrons from udsc-quark jets (upper-left), gluon jets (upper right) and b-quark jets (lower-left) for several parton energies in each case~\cite{part-pid1,part-pid2}. The measured FFs for light quarks and gluons are well described by model $\beta(u;p,q)$, shape parameters $(p,q)$ depending on parton species and energy scale. Dijet multiplicities are obtained as the best-fit coefficients of the unit-normal beta distribution. As expected, there is a substantial difference between quark and gluon FFs at larger jet energies, and a strong energy dependence of gluon jet shapes for smaller jet energies evident in the upper-right panel (the two solid curves correspond to $\sim$ 5 and 40 GeV gluons) ({\em cf.} Sec.~\ref{betasum}). The b-quark data in the lower-left panel are not well described by the beta distribution. The best-fit beta distributions for $\sqrt{s} = 91.2$ GeV udsc quark and $Q = 80.2$ GeV (equivalent dijet energy) gluon jets (data with the best statistics) are repeated as the dashed ($\beta_\text{q}$) and dash-dot ($\beta_\text{g}$) curves respectively in all three panels to provide references. 

 
In Fig.~\ref{fig12} (lower-right panel) we compare FFs from different parton types in a more differential format. As noted, $\beta_\text{q}$ (dashed curves) for light fragments in udsc jets and $\beta_\text{g}$ (dash-dot curves) for gluon jets are approximate limiting cases for all $\beta(u;p,q)$.  We therefore define $v_{max} \equiv \ln(\beta_\text{q} + \beta_\text{g})$, $v_{min} \equiv - \ln(1/\beta_\text{q} + 1/\beta_\text{g})$ and normalized variable $v(\beta) \equiv (\ln \beta - v_{min})/(v_{max} - v_{min})$, with $v(\beta_\text{q}) + v(\beta_\text{g}) = 1$. We plot $v(\beta_\text{q})$ (dashed), $v(\beta_\text{g})$ (dash-dot) and $v(\beta_\text{b})$ (solid) in the lower-right panel with the corresponding data for $\sqrt{s} = 91.2$ GeV quarks and $Q = 80.2$ GeV gluons (also {\em cf.} data and solid curves in Fig.~\ref{figfit}). 

\begin{figure}[h]
\includegraphics[width=3.3in,height=1.95in]{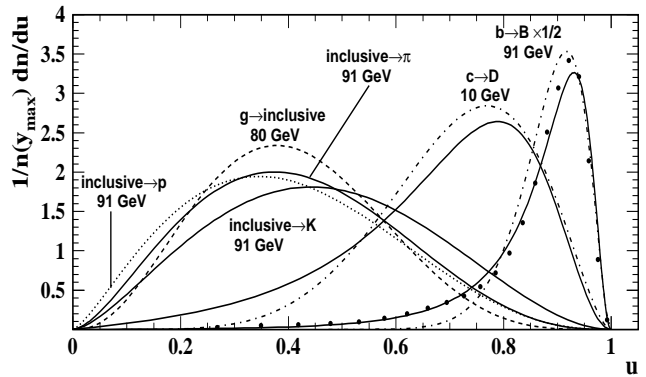}
\caption{\label{fig13}
Distributions on $u$ for several quark/meson flavor combinations, showing evolution of the $g(u,y_{max})$ shape with quark/meson mass. The c $\rightarrow$ D data are from~\cite{cleoD}, and the b $\rightarrow$ B data are from~\cite{alephB}. The energies are dijet energies. 
}
\end{figure}

The light-fragment distribution from b quarks (solid dots) coincides with $v(\beta_\text{q})$ (and open circles) for $u < 0.7$, but diverges sharply from the quark-jet trend above that point and descends towards $v(\beta_\text{g})$ (and open triangles) for $u > 0.7$. The b-quark fragment data were reduced by 10\% to coincide with the quark-jet curve below $u \sim 0.7$. The initial normalization is represented by the beta-distribution fit  $v(\beta_\text{b})$ (solid curve) with mode near 0.5. With this more differential format we confirm that b-quark light-hadron fragments are not well described by a beta distribution. The exceptional softness of the b-quark FF (for unidentified fragments) was anticipated theoretically~\cite{webber-hadro} (and {\em cf.} Fig.~\ref{fig13}).

In Fig.~\ref{fig13} we summarize FF data and models for several fragment and parton types. The pion, kaon and proton FFs are beta-distribution fits to 91 GeV identified-fragment data (pion and kaon data are shown in Fig.~\ref{fig10}). The gluon FF is the beta distribution defined at 80 GeV by $(p,q)$ systematics in Fig.~\ref{fig8} ({\em cf.} comparison with FF data in Fig.~\ref{figfit} -- right panel). The solid dots are b $\rightarrow$ B data from~\cite{alephB} compared to a best-fit beta distribution (dash-dot curve) and theory (solid curve). Low-statistics c $\rightarrow$ D data from~\cite{cleoD} are summarized by a best-fit beta distribution (dash-dot curve) and theory (solid curve).

The two solid curves on the right of Fig.~\ref{fig13} are from a theoretical treatment of heavy-quark fragmentation in which the FF for $Q \rightarrow H(Q\bar q) + q$ is approximated by $D_Q^H(x_p) \propto 1/\{1 - 1/x_p - \epsilon_Q / (1-x_p)\}^2/x_p$, with $\epsilon_Q \propto 1/m^2_Q$~\cite{heavy}. The agreement of $D_b^B(u)$ (right-most solid curve)  with b-quark data (solid points)~\cite{alephB} using $\epsilon_Q = 1.16/m^2_b = 0.055$ and $m_b \sim 4.6$ GeV/c$^2$ is good. The dash-dot curve is the best-fit beta distribution with $(p,q) = (23,3)$ which does not describe the $b \rightarrow B$ data. That failure may be related to the exceptional behavior of b $\rightarrow$ light hadrons discussed in connection with Fig.~\ref{fig12} (lower panels). The solid curve for $c \rightarrow D$ is $D_c^D(u)$ from the heavy-quark theory treatment, with $\epsilon_Q = 0.57/m^2_c =  0.29$ and $m_c \sim 1.4$ GeV/c$^2$. The associated dash-dot curve is a beta distribution with $(p,q) = (7.0,2.8)$ which best describes the data from~\cite{cleoD}. Both curves are consistent with the data, but the data errors are large below the FF peak mode. 



FF modes increase monotonically with increasing {meson} and parton mass. However, the proton FF mode for udsc jets is lower than the inclusive hadron mode for gluon jets and the FF is significantly broader. The kaon FF shows the effect of the heavier s-quark mass, consistent with the trend for charm and bottom quarks (however, see the next paragraph). The FF mass dependence on normalized rapidity $u$ is subtle compared to the kinematic dependence on meson and parton masses encountered on $p_t$, $\xi_p$ or $y$. 

To summarize flavor dependence, the beta distribution describes the FF data for identified light quarks and gluons fragmenting to identified light mesons or baryons, providing a compact representation of the flavor dependence of fragmentation. The quality of the description is not good for heavy quarks fragmenting to light or heavy mesons. However, the region near $u = 1$ can be compared with non-perturbative trends for light-quark fragmentation extrapolated to small $Q^2$ in Fig.~\ref{fig8} (right panel), where the fragmentation `cascade' is a single splitting or no splitting (parton $\rightarrow$ hadron).

\section{Fitting $\beta(u;p,q)$ to Data} \label{distparam}

We now fit the beta distribution to a sample of measured FFs falling in three groups: 1) the five fiducial FFs for unidentified fragments from flavor-inclusive partons distinguished by nearly complete coverage of the kinematically-allowed fragment momenta~\cite{frag,tasso} and a selection of data for 2) identified fragments and 3) identified partons to explore the role of hadron and parton species in fragmentation. Data in the form $D(u,y_{max})$ are fitted with model function $2n(y_{max})\, \beta(u;p,q)$ ({\em cf.} Sec.~\ref{beta0}), minimizing $\chi^2$ while freely varying parameters $2n,~p$, and $q$, with $u = (y - y_{min})/ (y_{max} - y_{min})$ and $y_{min}$ constrained to specified values based on systematics studies.

\subsection{Inclusive fragments from inclusive partons} \label{ymin}

We first fit FF data for inclusive hadrons and partons. Table~\ref{betapar1} contains the best-fit parameters for the five fiducial FFs (OPAL~\cite{frag} and TASSO~\cite{tasso} data) with $y_{min} = 0.35$ ($p = 0.05$ GeV/c). 
The model functions with starred energies are compared to data in Figs.~\ref{fig2} - \ref{fig5}. As noted, the FF shape is {\em nearly} independent of $\sqrt{s}$, but there is a significant trend for $q$ to increase and $p$ to decrease with increasing energy scale, shifting the FF mode to smaller $u$. The fitted multiplicities agree with the q-\=q multiplicity curves and data in Fig.~\ref{fig7}.


\begin{table}[h]
\caption{Unidentified fragments from unidentified partons: Beta-distribution parameters from $\chi^2$ fits to fragmentation functions for five energies.  The FFs for starred energies are plotted in Figs.~\ref{fig2} - \ref{fig5}.
\label{betapar1}
}
\begin{center}
\begin{tabular}{|c||c|c|c||c|} \hline
 $\sqrt{s}\, $(GeV) & 2n & p & q  & $\chi^2/\nu$ \\ \hline\hline
 14* & 8.8$\pm0.10$   &  2.95$\pm0.08$ &  3.52$\pm0.07$ & 16/18 \\ \hline
  22  &  10.7$\pm0.15$ & 2.91$\pm0.06$ & 3.52$\pm0.08$  &  25/20 \\ \hline
  35 & 13.4$\pm0.05$  & 2.84$\pm0.02$ & 3.50$\pm0.02$ &  148/22 \\ \hline 
 44*   & 14.6$\pm0.10$  &  2.89 $\pm0.03$ &  3.52$\pm0.04$ &  49/22 \\ \hline 
 91.2* & 20.4$\pm0.05$  &  2.84$\pm0.01$ &  3.67$\pm0.01$  &  86/51 \\ \hline 
\end{tabular}
\end{center}
\end{table}


\begin{figure}[h]
\mbox{
\includegraphics[width=1.65in,height=1.65in]{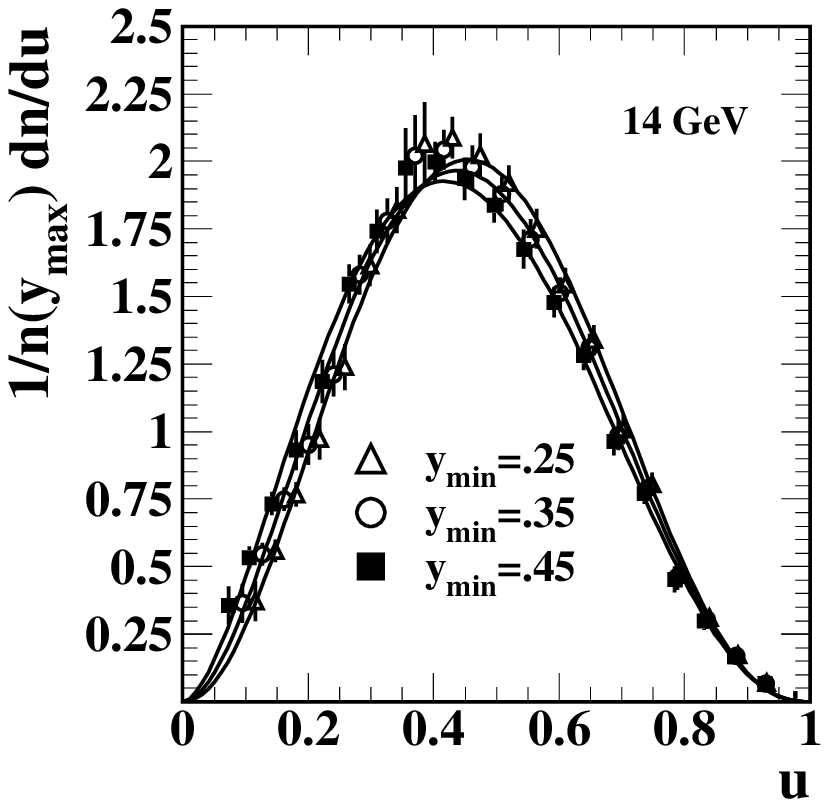}
\includegraphics[width=1.65in,height=1.65in]{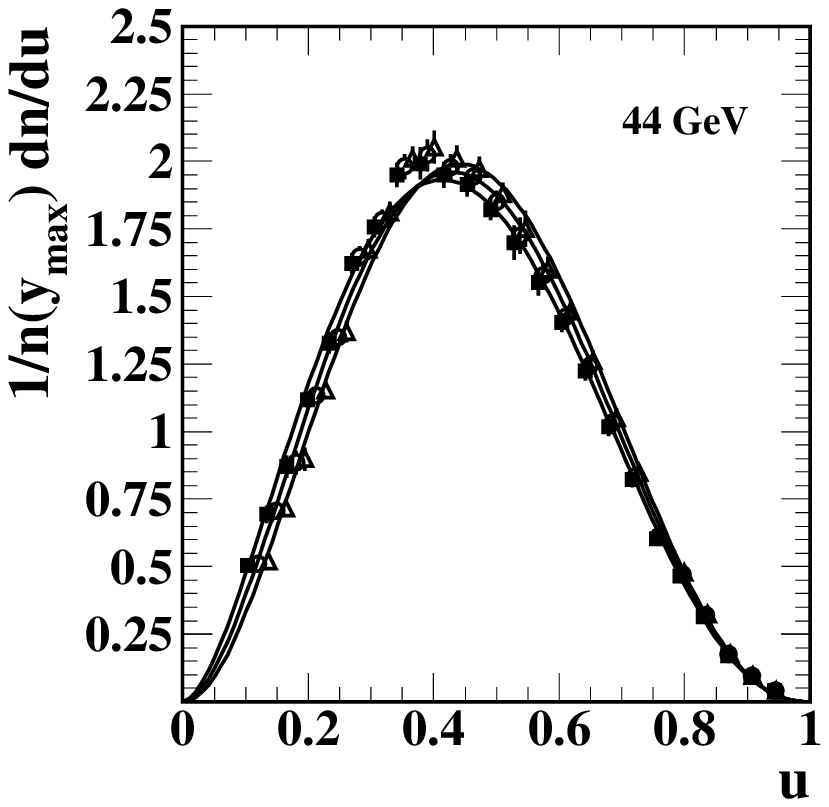}
}
\mbox{
\includegraphics[width=1.65in,height=1.65in]{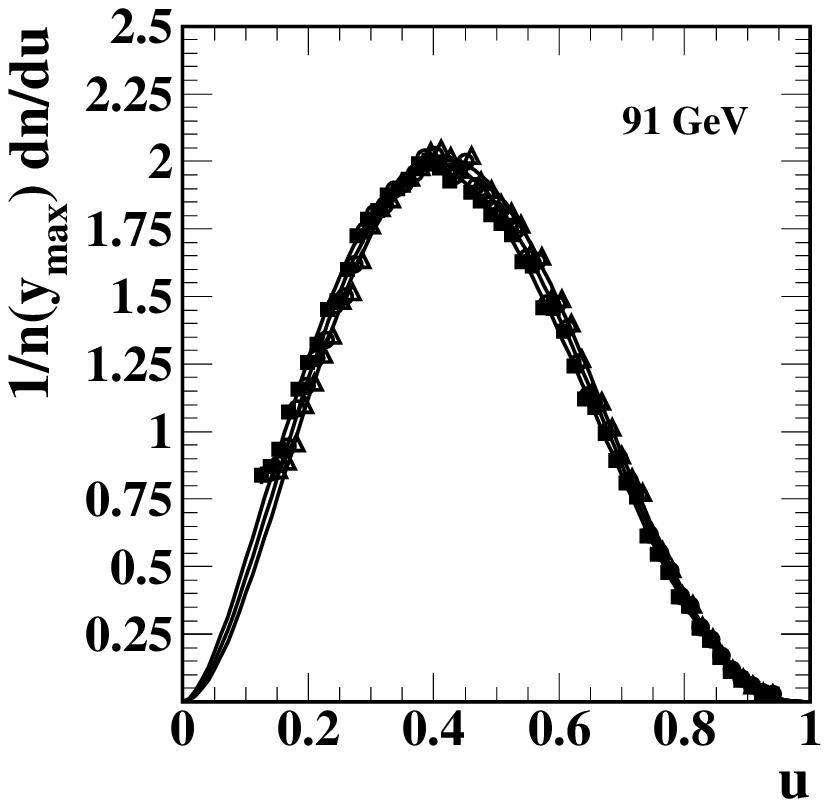}
\includegraphics[width=1.65in,height=1.65in]{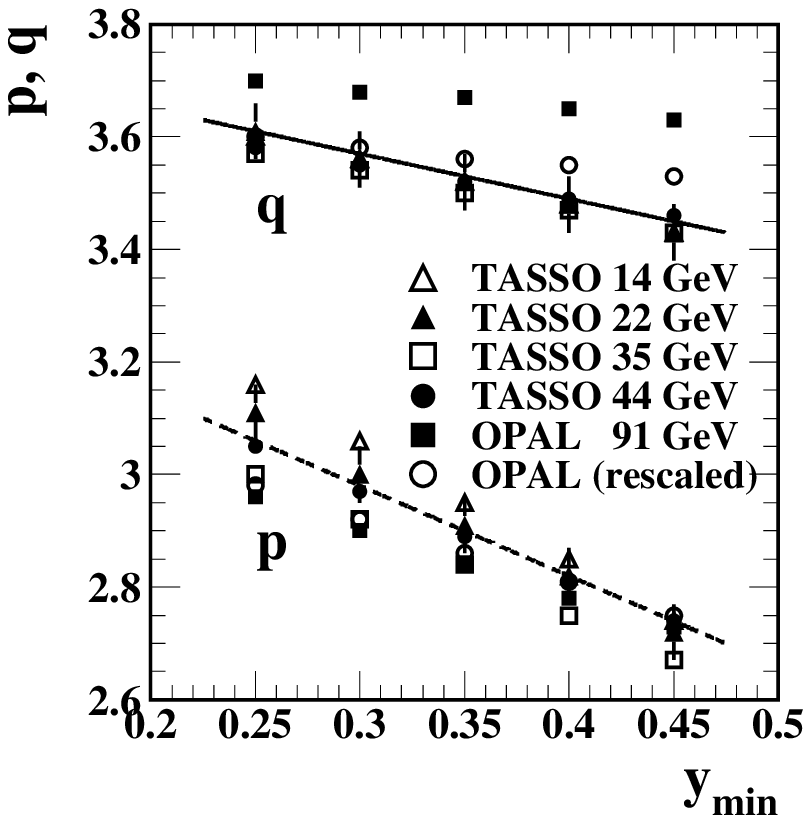}
}
\caption{\label{figymin}
A study of the systematic effect on fiducial FFs and their beta parameters $(p,q)$ of variations with $y_{min}$. The low ends of the FFs are most sensitive, and therefore p and FFs for low $y_{max}$ are affected most. The trends in the lower-right panel may be compared with the $(p,q)$ entries in Table~\ref{betapar1}. 
}
\end{figure}

Fig.~\ref{figymin} shows the systematic dependence of beta parameters $(p,q)$ on the choice of $y_{min}$. The first three panels illustrate the variation of fits and data on $u$ with $y_{min}$ for energies 14, 44 and 91 GeV.  $\chi^2$ variations are small over the interval shown. Variation with $y_{min}$ is greater for smaller $u$ and smaller $y_{max}$. Those trends are reflected in the summary of $(p,q)$ variations in the lower-right panel: the $p$ variation is greater, and more so for lower energy. The lines have slopes 0.8 (solid) and 1.6 (dashed). We set $y_{min} = 0.35$ for all inclusive fits and discuss the related systematic uncertainties in Sec.~\ref{betasum}. The shift of OPAL (rescaled) $q$ data in the lower-right panel (closed squares to open circles) results from increasing all particle momenta by 6.5\% to test the effect of uncertainty in the momentum calibration. The $p$ data are much less affected.

\subsection{Identified fragments from inclusive partons}

We next explore the role of hadron identity in fragmentation, with $m_0 \rightarrow m_\text{hadron}$ assigned for both fragment and parton rapidities. Fits to identified hadron fragments from flavor-inclusive partons at $\sqrt{s} =$ 10 GeV~\cite{pid-10gev} and 91 GeV~\cite{pid-91gev} plotted in Fig.~\ref{fig10} and Fig.~\ref{fig11} (left panel) are presented in Table~\ref{betapar2}. Trends for fragmentation to light hadrons are summarized in Fig.~\ref{fig13}. For pions, parameter $p$ is smaller and $q$ larger than for inclusive hadrons in Table~\ref{betapar1}, shifting the peak mode to smaller $u$ as noted previously and understood as an effect of misidentifying kaons and protons as pions in the inclusive hadron fragment mixture. The pion fit $\chi^2$ is large; however the fit residuals are generally point-to-point random and substantially larger than the stated errors, especially toward the ends of the distribution. 

\begin{table}[h]
\caption{Identified fragments from unidentified partons: Beta-distribution parameters from $\chi^2$ fits to fragmentation functions for pions, kaons and protons at $\sqrt{s} =$ 8 - 18 and 91.2 GeV
\label{betapar2}
}
\begin{center}
\begin{tabular}{|c||c|c|c|c||c|} \hline
 FID & 2n & p & q & $y_{min}$ & $\chi^2/\nu$ \\ \hline\hline
\multicolumn{6}{|c|}{ $\sqrt{s} =$ 8 - 18 GeV} \\ \hline
$\pi^\pm$ & 5.63$\pm0.02$   &   2.92$\pm0.03$ &  3.96$\pm0.05$ &  0.35 &  89/49 \\ \hline
 K$^\pm$ & 0.88$\pm0.025$   &  3.15$\pm0.12$ &  3.84$\pm0.19$ &  0.10 &  33/39 \\ \hline
 p, \=p & 0.18$\pm0.02$   &  2.60$\pm0.40$ &  4.30$\pm1.00$ &  0.05 &  21/24 \\ \hline\hline
\multicolumn{6}{|c|}{ $\sqrt{s} =$ 91.2 GeV} \\ \hline
 incl. & 20.4$\pm0.05$   &  2.84$\pm0.01$ &  3.67$\pm0.01$ &  0.35 & 86/51 \\ \hline
 $\pi^\pm$ & 17.36$\pm0.03$   &   2.66$\pm0.01$ &  3.77$\pm0.01$ &  0.35 &  483/36 \\ \hline
 K$^\pm$ & 2.39$\pm0.03$   &  2.58$\pm0.03$ &  2.99$\pm0.01$ &  0.10 &  10/26 \\ \hline
 p, \=p & 1.10$\pm0.02$   &  2.36$\pm0.04$ &  3.58$\pm0.07$ &  0.05 &  17/23 \\ \hline
\end{tabular}
\end{center}
\end{table}

The 10 GeV kaon peak is similar to the pion peak. However, the 91 GeV kaon peak is much wider ($p+q$ is reduced) and the mode is shifted substantially to the right ($q - p$ is reduced) relative to the pion peak, consistent with the quark mass-dependence trend in Fig.~\ref{fig13}. The proton peak mode is shifted further to the left, beyond the pion and gluon peaks as shown in Fig.~\ref{fig13}, mainly by reduction of $p$. We note in passing that $y_{min} \sim \ln \left[\frac{m_0 + 50 ~\text{MeV/c}^2}{m_0}\right]$.

\subsection{Inclusive fragments from identified partons}

\begin{figure}[h]
\mbox{
\includegraphics[width=3.3in,height=1.75in]{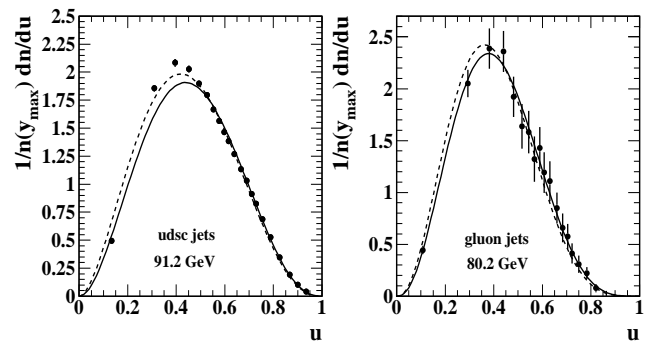}
}
\caption{\label{figfit}
Details of beta distribution fits to fragmentation functions for identified partons fragmenting to unidentified hadrons. The solid curves are free fits to data. The dashed curves are determined by the (p,q) systematics in Fig.~\ref{fig8} (left panel). 
}
\end{figure}

Finally, we consider data for unidentified hadron fragments from identified partons for two parton classes shown in Fig.~\ref{fig12}: udsc quarks (in combination) and gluons. The fit results are shown in Fig.~\ref{figfit} and Table~\ref{betapar3}. The `inclusive' table entry (first row) repeats the 91 GeV results from unidentified hadrons in Table~\ref{betapar1} for reference. Parameters for the free $\chi^2$ fit to the udsc FF data in the second row of the table reflect a width similar to the inclusive data ($q + p$ is similar), but the mode is shifted to slightly larger $u$ ($q - p$ is smaller). Details of the fitting procedure are shown in Fig.~\ref{figfit} (left panel). The points are substantially larger than the reported errors. The free fit (solid curve) is strongly influenced by the single point at $u \sim 0.13$. The $\chi^2$ is large, and the fit function misses the data near the peak. As for the pion fragment FF data the udsc data errors appear to be underestimated. 




\begin{table}[h]
\caption{Unidentified fragments from identified partons: Beta-distribution parameters from $\chi^2$ fits to fragmentation functions for udsc quarks with $\sqrt{s} = 91.2$ GeV and gluons with $Q = 80.2$ GeV.
\label{betapar3}
}
\begin{center}
\begin{tabular}{|c||c|c|c|c|c|} \hline
 PID & 2n & p & q & $y_{min}$ & $\chi^2/\nu$ \\ \hline\hline
 incl. & 20.4$\pm0.05$   &  2.84$\pm0.01$ &  3.67$\pm0.01$ &  0.35 & 86/51 \\ \hline
udsc & 18.36$\pm0.04$   &  2.99$\pm0.015$ &  3.55$\pm0.015$ &  0.35 & 209/19 \\ \hline
udsc & param. $(p,q)$   &  2.85$\pm0.05$ &  3.58$\pm0.05$ &  0.35 & 550/19 \\ \hline \hline
 gluon & 27.2$\pm0.4$   &  3.50$\pm0.10$ &  5.10$\pm0.15$ &  0.35 & 5.7/22 \\ \hline
 gluon &  param. $(p,q)$  &  3.43$\pm0.10$ &  5.30$\pm0.10$ &  0.35 & 14.4/22 \\ \hline
\end{tabular}
\end{center}
\end{table}

The dashed curve is constrained by the $(p,q)$ energy systematics in Fig.~\ref{fig8} (left panel) consistent with fits to the fiducial {\em inclusive} FFs (second udsc row of the table). The peak of the data FF is better described, but the increased deviation from the small-$u$ point greatly increases the $\chi^2$. We expect the udsc FF to differ slightly from the flavor-inclusive FF (small shift to the right) due to the absence of gluon fragments, and the free fit is consistent with that expectation. We expect the udsc/inclusive multiplicity ratio to be 0.94~\cite{opal-mult} but observe 18.36/20.4 = 0.90, suggesting that the udsc FF height is underestimated by the free fit. 





In Fig.~\ref{figfit} (right panel) we show a free fit to FF data from gluon jets (solid curve) producing the fit parameters in the first gluon row of Table~\ref{betapar3}, which are plotted as open squares in Fig.~\ref{fig8} (left panel) and provide constraints on the gluon $(p,q)$ energy systematics discussed in Sec.~\ref{betasum}. The free fit has an unusually small $\chi^2$; the data errors above the mode seem large compared to the residuals there. The dashed curve is a `fit' with parameters constrained to the 80 GeV $(p_\text{g},q_\text{g})$ systematic values from Fig.~\ref{fig8} (left panel) and reported with its $\chi^2$ value in the second gluon row of Table~\ref{betapar3}.

\section{Energy Scale Dependence}

We can combine fits to fiducial FF data and dijet multiplicity data to determine the energy dependence of $(p,q)$ for quark and gluon jets over a broad energy range. Fits to data $g(u,y_{max})$ with model $\beta(u;p,q)$ determine specific values $(p,q)$ which constrain parameterized curves $(p(y_{max}),q(y_{max}))$. Fits to $2n(y_{max})$ data {\em via} the $\langle x_E \rangle$ integral of $\beta(u;p,q)$ also constrain the parameterizations, especially important in energy intervals where there are no FF data available. The resulting $(p,q)$ energy trends efficiently represent $e^+$-$e^-$ FFs over a broad energy range and provide a basis for extrapolating FFs to low $Q^2$.

\subsection{Energy conservation sum rule}

The total FF $D(x_E,s) = \sum_h D^h(x_E,s)$ (sum over all hadron species) integrates to {total} dijet multiplicity $\int_{2m_0/\sqrt{s}}^1 dx_E\, D(x_E,s) = 2n_{tot}(s)$ and satisfies the energy sum rule ($ESR$) $\int_{2m_0/\sqrt{s}}^1 dx_E\, x_E\,  D(x_E,s) = 2$~\cite{biebel}. The ratio of the integrals defines mean energy fraction $\langle x_E \rangle = 1/n_{tot}(s)$. Switching to $(u,y,y_{max})$, since $D(u,y_{max}) \equiv 2n_{tot}(y_{max})\, g(u,y_{max})$ and $g(u,y_{max}) \approx \beta(u;p,q)$ we have $\langle x_E \rangle \approx \int_{0}^1 du\, x_E(u,y_{max})\,  \beta(u;p,q)$, with $x_E(u,y_{max}) = \cosh[y(u)] / \cosh(y_{max})$ and $y(u) = u\, y_{max} + (1-u)\, y_{min}$. Those relations connecting $\beta(u;p,q)$ to $n_{tot}(y_{max})$ are used below to obtain the energy dependence of $(p,q)$ from multiplicity data.

Given several hadron species $h$ with FFs $D^h(x_E)$ and dijet multiplicities $2n_h$ we expect $ESR = \sum_h 2n_h\, \langle x_E \rangle_h = 2$, provided {\em all} species are integrated. However, if only charged hadrons are detected we expect $ESR \sim 2/3 \times 2 \sim 1.33$. We can test the charged-fraction $ESR$ using the fits to charged pion, kaon and proton data at 91 GeV from Table~\ref{betapar2}. The $ESR$ for inclusive FFs can be tested with the fits from Table~\ref{betapar1}. In general, if $f$ is the ESR fraction for detected particles ($f \sim 2/3$ for the charged-hadron fraction) we expect the relation $\langle x_E \rangle = f/n(y_{max})$ between monojet (charged-particle) multiplicity and energy fraction, which we use below to relate multiplicities to beta parameters $(p,q)$.  

\begin{figure}[h]
\mbox{
\includegraphics[width=3.3in,height=1.75in]{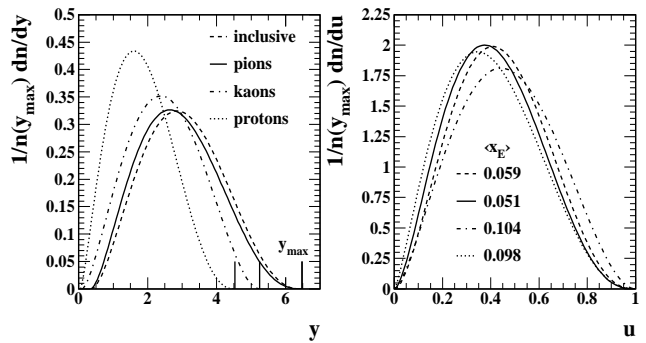}
}
\caption{\label{engint}
Unit-normal model functions $\beta(y~\text{or}~u;p,q)$ from fits to FFs for identified fragments (pions, kaons and protons) and inclusive fragments, all from inclusive partons at $\sqrt{s} = 91.2$ GeV, plotted on rapidity $y$ and normalized rapidity $u$. 
LONGER LINES
}
\end{figure}

Beta distribution fits to FFs $g^h(y ~\text{or}~ u,y_{max})$ for identified pions, kaons and protons at 91.2 GeV from Table~\ref{betapar2} and the FF for inclusive hadrons are plotted on rapidity $y$ and normalized rapidity $u$ in Fig.~\ref{engint}. The parameters for those curves are used to obtain $\langle x_E \rangle_h$ for each hadron species using the correct hadron mass and $\langle x_E \rangle_{incl}$ for the inclusive distribution assigning the pion mass to all hadrons. We use the $2n$ fit values in the tables to obtain $\sum_h 2n_h\, \langle x_E \rangle_h = 1.25\pm0.03$ ($f = 0.62$) and  $ 2n_{incl}\, \langle x_E \rangle_{incl} = 1.18\pm0.05$ ($f = 0.59$) for identified and inclusive charged fragments. If $\langle x_E \rangle_{incl}$ is calculated with the weighted-mean mass 0.2 GeV (weighted by the hadron multiplicities in Table~\ref{betapar2}) we obtain $ESR = 1.4\pm0.05$. The same procedure applied to the fits to lower-energy data from Table~\ref{betapar2} gives $ESR \sim 1.1$. The exact energy scale for the lower-energy sum rule is not clear because of the scale range, but the result is roughly consistent with expectations. For the inclusive analysis below assuming the pion mass we use $ESR$ factor 1.18.




\subsection{Dijet multiplicities from  $\beta(u;p,q)$ shapes} \label{jetmult}

Dijet multiplicity 2$n$ can be obtained directly by integrating measured and extrapolated FF data, as in Tables~\ref{betapar1} - \ref{betapar3}. However, as we have just shown there is a correspondence between $2n(y_{max})$ and the {\em shape} of data FF $g(u;y_{max})$ or fitted model function $\beta(u;p,q)$ determined by parameters $[p(y_{max}),q(y_{max})]$. We have obtained for inclusive charged fragments with pion mass assignment the relation $2n(y_{max}) = 1.18 / \int_{0}^1 du\, x_E(u,y_{max}) \beta(u;p,q)$ at 91.2 GeV which we now use to relate energy trends of FF shape parameters $(p,q)$ to fragment multiplicities. Measured multiplicity trends on parton energy thereby provide constraints on the energy dependence of FF parameters $(p,q)$, even in energy intervals where there are no measurements of FFs. 









\begin{figure}[h]
\includegraphics[width=3.3in,height=1.86in]{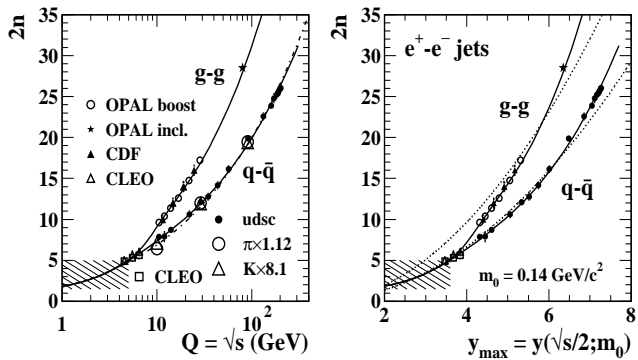}
\caption{\label{fig7}
Dijet charged-particle multiplicity {\em vs} energy scale $Q$ (dijet energy) plotted in a conventional format (left panel) and {\em vs} parton rapidity assuming the pion mass (right panel). The solid curves are quark and gluon dijet multiplicities $2n_\text{q}$ and $2n_\text{g}$ obtained from the $(p,q)$ parameterizations in Fig.~\ref{fig8} (left panel). The dash-dot curve in the left panel is from a 3NLO pQCD expression. 
The udsc quark-jet multiplicities for unidentified hadrons (solid dots) are taken from a survey in~\cite{lep-mult2}. The dotted curves in the right panel illustrate quadratic trends $A\,(y_{max} - y_{min})^2$ (see text). 
}
\end{figure}

Fig.~\ref{fig7} shows dijet multiplicities $2n$ for g-g and q-\=q parton pairs. Precise multiplicity data for quark jets from two-jet events have been available for some time. New methods have produced similarly precise gluon-jet multiplicities from three-jet $e^+$-$e^-$ events. Data for gluon jets were obtained from CDF (closed triangles)~\cite{cdf-glue}, CLEO (open triangles)~\cite{cleo}, OPAL `jet-boost' algorithm (open circles)~\cite{opal-boost} and OPAL inclusive (star)~\cite{opal-sing}. Data for quark jets were obtained from a compilation (Table 6 in~\cite{lep-mult2}) and multiplied by factor 0.94 (the fraction of udsc jets in a flavor-inclusive sample~\cite{opal-mult}) to compare with the gluon jet multiplicities. The large points labeled $\pi$ and K are multiplicities from fits to identified fragment data~\cite{pid-10gev,pid-29gev,pid-91gev} in Table~\ref{betapar2} plotted with the indicated multipliers. The hatched regions represent the domain of low-$Q^2$ partons which motivates this extrapolation study.

The solid curves in Fig.~\ref{fig7} are multiplicities derived from the $(p,q)$ energy trends using the relations defined above. The $(p,q)$ parameterizations are adjusted to fit the multiplicity data but constrained by $(p,q)$ values from fits to fiducial FFs. The resulting $(p,q)$ energy dependence is described in the next subsection. Because $2n \propto 1/ \langle x_E \rangle$ and $\langle x_E \rangle$ is monotonic with mean $\bar u = p / (p + q)$, multiplicities are mainly determined by ratio $q/p$ or difference $q - p$ ({\em i.e.,} the mode or mean of the beta distribution), and only weakly dependent on sum $q + p$ (the width). A unique description of $(p,q)$ over a broad energy range requires fits to multiplicity trends supplemented by the fits to fiducial FFs described in the previous section. 

Quark-jet multiplicities are described in the MLLA by 3NLO expression $n_\text{q}(Y) = K/2.25\cdot Y^{-a_1\,C^2}\, \exp\{ 2C\,\sqrt{Y} + a\, \delta(Y)\}$, with $Y = \ln(\sqrt{s}/\Lambda)$, $C = \sqrt{4n_c/b}$ and $b = (11 n_c - 2 n_f)/3$~\cite{lep-mult3}. We used $a_1 = 0.3$ from~\cite{lep-mult4} and $K = 0.13$ and $\Lambda = 0.15$ GeV from Table 1 and the functional form of $\delta_g(Y)$ from Fig. 3 in~\cite{lep-mult3} for $\delta(Y)$ (all for $n_f = 5$). We set the coefficient of $\delta (Y)$ to $a = 1.8$ to obtain the best agreement with quark-jet data, shown by the dash-dot curve in Fig.~\ref{fig7} (left panel) just visible relative to our parameterization (solid curve).

Variation of dijet multiplicities in the form $A\,(y_{max} - y_{min})^2$ would be expected for the self-similar scaling illustrated in Fig.~\ref{fig1} with fixed FF mode $u^*$. Quadratic trends for quark and gluon jets are illustrated by the dotted curves in Fig.~\ref{fig7} (right panel), with $A = 0.5$ for quarks 
and $1.45\cdot 0.5$ for gluons.  Deviations from the quadratic trend for quark-jet multiplicities in Fig.~\ref{fig7} correspond to the linear variation of $(p_\text{q},q_\text{q})$ with $y_{max}$ above $y_{max} = 4.5$ in Fig.~\ref{fig8} (left panel) which shift $u^*$ to smaller values, as illustrated in the right panel of that figure. Gluon-jet multiplicities deviate more dramatically from the quadratic trend at lower energies, moving from the quark-jet curve to a gluon-jet trend about 50\% larger within the energy interval $y_{max} = 3.5 - 5$ ($Q = 5 - 20$ GeV) as the quark-gluon color charge difference emerges. Above 20 GeV the gluon-jet multiplicities reflect the smaller linear variation of $(p_\text{g},q_\text{g})$ with $y_{max}$ in that energy interval.

\subsection{Energy dependence of $\beta(u;p,q)$ parameters} \label{betasum}

\begin{figure}[h]
\includegraphics[width=1.65in,height=1.82in]{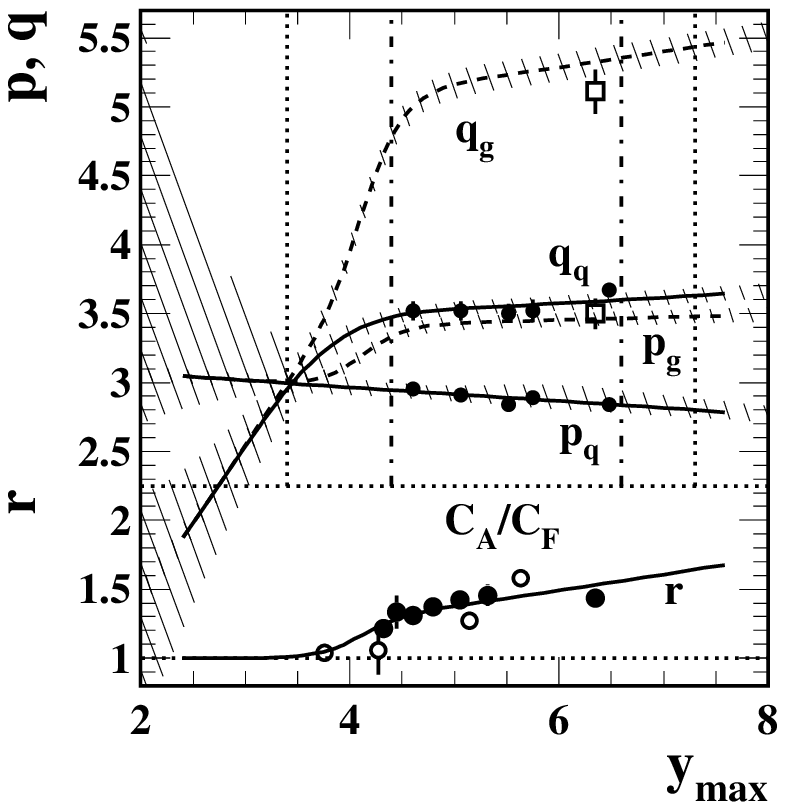}
\includegraphics[width=1.65in,height=1.84in]{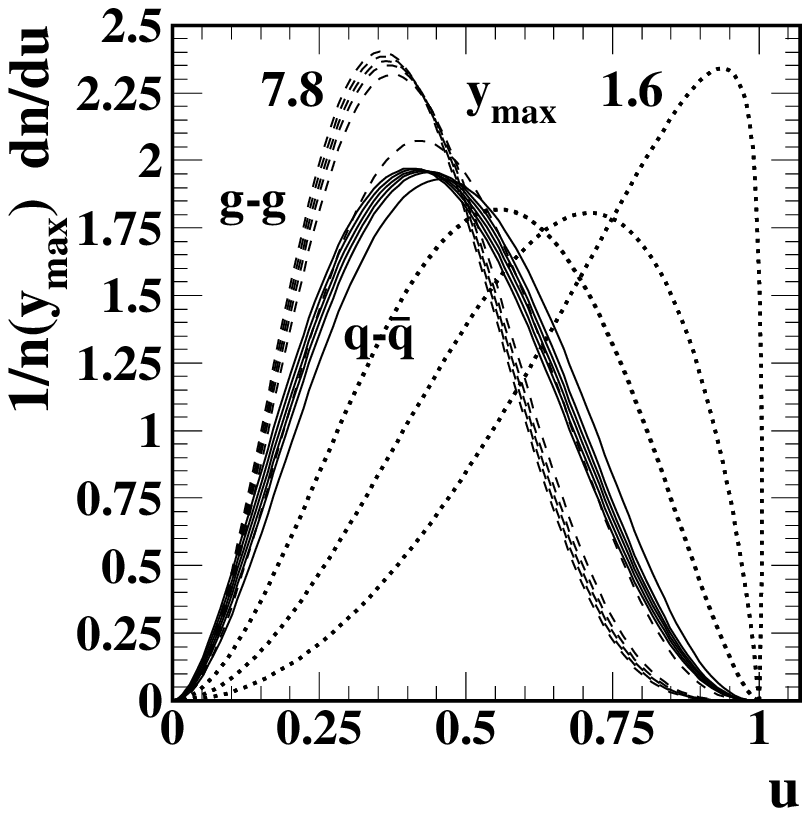}
\caption{\label{fig8} 
Left panel: Beta-distribution parameters $(p_\text{q},q_\text{q})$ and $(q_\text{g},p_\text{g})$ respectively for light-quark (solid) and gluon (dashed) jets and corresponding gluon-to-quark-jet multiplicity ratio $r$ {\em vs} parton rapidity $y_{max}$.
Right panel: Unit-normal FFs (beta distributions) obtained from parameters in the left panel plotted on normalized rapidity $u$ for quark (solid) and gluon (dashed) jets and for nine equal-spaced values of parton rapidity $y_{max}$ illustrating peak shape evolution with energy scale. 
}
\end{figure}

Fig.~\ref{fig8} (left panel) shows the $(p,q)$ energy dependence which produces the quark and gluon jet multiplicities (solid curves) in Fig.~\ref{fig7} and the solid curves compared to fiducial FFs in Figs.~\ref{fig2} - \ref{fig5}. Those curves summarize the energy dependence of udsc and gluon fragmentation to unidentified hadrons in $e^+$-$e^-$ collisions. We assume that the shapes of inclusive (dominated by light quarks) and udsc FFs are approximately the same, as in Fig. 6 of~\cite{part-pid1}. Inclusive and udsc jet FFs at 91.2 GeV are compared in Table~\ref{betapar3} and Fig.~\ref{figfit} (left panel) and found to be similar. The vertical dotted lines mark the limits of multiplicity measurements, while the vertical dash-dot lines mark the limits of measured FFs used in this analysis. The upper ten solid points represent the fiducial FFs. The open squares represent the single gluon FF in Fig.~\ref{figfit} (right panel) which constrains $(p_\text{g},q_\text{g})$.

The $(p,q)$ curves in Fig.~\ref{fig8} (left panel) are described by 
\bea \label{ppp}
p_\text{q} &=& 2.90\pm0.05 - (0.05\pm0.01)\,(y_{max} - 5.3)  \\ \nonumber
q_\text{q} &=&  3.50\pm0.05 + (0.05\pm0.01)\,(y_{max} - 5.3)  \\ \nonumber
&  -& (0.8\pm 0.2)\, (y_{max} - 4.5) 
 [\tanh(y_{max} - 3.5)-1]/2 \\ \nonumber
p_\text{g} &=& p_\text{q} +  [(0.07\pm0.02)\,(y_{max} - 5.3) + 0.55\pm0.05] \times \\ \nonumber
 & &\{ \tanh[(2.5\pm0.5)\, (y_{max}-4.1\pm0.1)] + 1\}/2 \\ \nonumber
q_\text{g} &= & q_\text{q} + [(0.07\pm0.02)\,(y_{max} - 5.3) + 1.70\pm0.07] \times \\ \nonumber
 & &\{ \tanh[(2.5\pm0.5)\, (y_{max}-4.1\pm0.1)] + 1\}/2  \\ \nonumber
\eea
The $p_\text{q}$ expression in Eq.~(\ref{ppp}) is determined only by linear interpolation and extrapolation of fits to the hadron- and parton-inclusive fiducial FFs. Given that definition of $p_\text{q}$ the curve for $q_\text{q}$ is then defined only by the fit to the udsc quark-jet multiplicity data in Fig.~\ref{fig7}. The agreement in Fig.~\ref{fig8} between $q_\text{q}$ determined by fitting light-quark multiplicities $2n_\text{q}$ (upper solid curve) and by fitting individual inclusive-parton FFs (upper solid points) indicates the consistency of the two methods. The  expression for $p_\text{g}$ is guided by the fit to a single gluon FF denoted by the lower open square point in Fig.~\ref{fig8} (left panel), but is also influenced by its impact on Fig.~\ref{fig17}. The expression for $q_\text{g}$ is then determined relative to $p_\text{g}$ mainly by gluon multiplicities $2n_\text{g}$, but is also influenced by its impact on Fig.~\ref{fig17}. See the discussion of that figure for more details.

The error bands in the central region represent correlated systematic errors related to the uncertainty in $y_{min}$. As $y_{min}$ varies the dominant effect is common displacement of $p$ and $q$ ({\em cf.} Fig.~\ref{figymin}). Multiplicity depends mainly on the mode or mean of the fragment distribution, and therefore primarily on the ratio or difference of $p$ and $q$. Multiplicity is therefore insensitive to the choice of $y_{min}$. The FF width on the other hand depends directly on the sum $p+q$ and is therefore more influenced by the choice of $y_{min}$.
To the right of the left dash-dot line the $(p,q)$ vary slowly and linearly with increasing energy scale. The energy dependence for light quarks implies a slight reduction of the mode with the peak width unchanged, consistent with the fiducial FFs in this study ({\em e.g.,} Fig.~\ref{fig5}). The gluon FF shows similar mode variation, but the width is also reduced with increasing energy. 

Below the left dash-dot line ($Q \sim 10$ GeV) the $(p,q)$ change rapidly. The multiplicity data, especially the CLEO data, require a sharp drop in $q$ in that energy interval for both quarks and gluons which is effected by the $\tanh$ term in $q_\text{q}$ of Eq.~(\ref{ppp}). The convergence of the quark and gluon $(p,q)$ at the energy scale defined by the lower dotted line, again required by the CLEO data, is effected by the $\tanh$ terms in $p_\text{g}$ and $q_\text{g}$. 
Below $y_{max} = 3.6$ ($Q  =5$ GeV) there is no guidance from data, but we speculate as follows. At 5 GeV the average jet multiplicity is $\sim 2.5$ and there is no distinction between quark and gluon jets, $p \sim q$ and the FF is therefore symmetric about the midpoint on $u$. We argue that at lower energies the mean jet multiplicity approaches one and the FF approaches a delta function at $u = 1$, requiring $q \rightarrow 1$ and $p \rightarrow \infty$. We sketch those trends with large error bands in the left panel as a simple extrapolation of the trends derived from data. 

In Fig.~\ref{fig8} (right panel) we show a sequence of model functions for nine equal $y_{max}$ steps from 1.6 to 7.8, with parameters derived from the $(p,q)$ curves in the left panel for gluon and quark FFs. The modes for quark jets (solid curves) and gluon jets (dashed curves) move from left to right with decreasing energy scale, and the dotted curves for $y_{max} < 3.6$ ($Q < 5$ GeV) represent both parton types in common.  Below 5 GeV the FFs slew to the right and may approach a delta-function limit at $u  =1$. Those low-$Q^2$ trends can be compared with the theoretical description of heavy quark fragmentation on the right of Fig.~\ref{fig13}. 


The energy dependence of gluon-to-quark-jet multiplicity ratio $r = n_\text{g} / n_\text{q}$ derived from the beta parameters is plotted as the lowest solid curve in Fig.~\ref{fig8} (left panel). The open points are taken from~\cite{cleo2} and the solid points are from~\cite{delphi-scalviol}. $r$ is expected to approach the ratio of color factors $C_A/C_F = 2.25$ at large $Q$. The ratio indeed increases monotonically with $y_{max}$ from unity at $y_{max} \sim 3.5$ ($p \sim$ 2 GeV/c and $n \sim 2$), but the approach to $C_A/C_F$ is slow. The overall trend is in rough agreement with theory~\cite{eden,lep-mult4}. 

The ratio slope $r' = dr/dy_{max} \sim dr/d\ln(Q)$ is also of theoretical interest but difficult to calculate since it is very sensitive to perturbative corrections~\cite{lep-mult4}. Slope $r'$ derived from $(p,q)$ rises to peak value $0.45$ at $y_{max} \sim 4$, then falls to 0.11 at $y_{max} = 5$ ($Q =$ 20 GeV) and rises linearly to 0.13 at  $y_{max} = 8$ ($Q = $ 400 GeV). The value r' = 0.1 is in rough agreement with theory~\cite{lep-mult4}. In contrast to the slow evolution of $r$ the rapid separation of $q_\text{q}$ and $q_\text{g}$ with increasing energy between the lower dotted and dash-dot vertical lines of Fig.~\ref{fig8} (left panel) contrasted with a nearly fixed difference between them above that region may provide a clearer manifestation of the emergence of color charge.



\subsection{Fragmentation functions on $(y,y_{max})$} \label{fragyymax}

We can use the parameterized beta distribution to construct a 2D fragment distribution on $(y,y_{max})$ as follows. Form factor $\beta(u;p,q)$ describes the shapes of FFs over a broad $Q^2$ interval. The beta distribution in turn determines multiplicity $n(y_{max})$ through $\langle x_E \rangle$ over the same range (Fig.~\ref{fig7}). We combine the two factors to form $D(y,y_{max}) = 2n(y_{max})\, \beta[u(y,y_{min},y_{max});p(y_{max}),q(y_{max})]$.

\begin{figure}[h]
\includegraphics[keepaspectratio,width=3.3in]{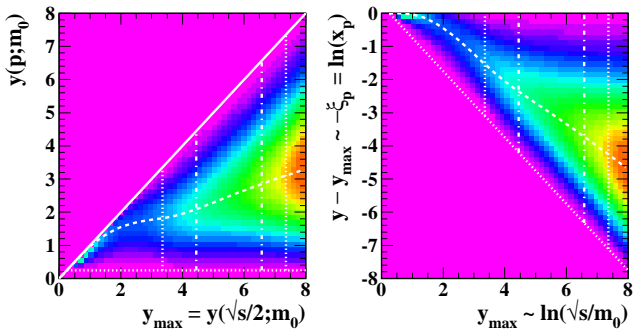}
\caption{\label{fig14}
Left panel: Joint fragment distribution $D(y,y_{max})$ on fragment and parton rapidities for inclusive partons ($\sim$ udsc quarks) and inclusive hadrons. Fragmentation functions are vertical slices (conditional distributions) from the joint distribution. 
Right panel: The same distribution transformed to $(y - y_{max},y_{max})$, with $y_{max} - y \sim \xi_p$, the logarithmic relative momentum. The vertical dash-dot lines define the interval determined by fiducial FFs plus dijet multiplicities. The intervals between vertical dash-dot and dotted lines are defined only by multiplicity trends. The upper-right region of the right panel illustrates scaling violations. 
}
\end{figure}



In Fig.~\ref{fig14} (left panel) we plot $D(y,y_{max})$. The vertical dotted and dash-dot lines mark the same energies as in Fig.~\ref{fig8} (left panel). The dashed curve is a `locus of modes' (positions of maxima) of conditional distributions on $y$ for fixed $y_{max}$. The approach of that curve to the solid diagonal line ($y = y_{max}$) at lower left corresponds to the approach of the dotted curves in Fig.~\ref{fig8} (right panel) to $u = 1$. The horizontal dotted line denotes $y_{min}$, and 
$y_{max} = 8$ corresponds to $\sqrt{s} \sim 400$ GeV. This joint fragment density provides the basis for extrapolating FFs down to $Q \sim 1$ GeV ($y_{max} \sim 2$). Fig.~\ref{fig14} (right panel) is a transformation of the left panel onto $(y_{max},y-y_{max})$, with $y-y_{max} \sim \ln x_p = -\xi_p$, which illustrates in the upper-right corner scaling violations: variation of the fragment density with increasing $\ln(Q/\Lambda) \rightarrow y_{max}$ at constant $x_p$ or $\xi_p$ (constant $y - y_{max}$) ({\em cf.} Sec.~\ref{scalviol}).

\section{Peak statistics and pQCD} \label{ximode}

We have constructed a simple parameterized model of FFs for $e^+$-$e^-$ collisions which compares well with data. In this section we compare the energy dependence of peak statistics on $u$ and $\xi_p$ inferred from our parameterization with predictions from pQCD. Fragmentation-function peak statistics predicted by pQCD~\cite{
one-two-particle,biebel,fw-peakstats,
cacf-delph} can be compared to peak statistics $u^*$ (mode), $\bar u$ (mean) and $\sigma^2_u$ (variance) for distribution $\beta(u;p,q)$. The mode for the $\beta$ distribution is $u^* = \frac{p-1}{p+q-2}$ and the mean is $\bar u = \frac{p}{p+q}$, with $(p,q)$ determined by the parameterizations in Fig.~\ref{fig8}. The {mode} on $y$ is $y^* = u^*\,y_{max} + (1-u^*)\,y_{min}$.

\begin{figure}[h]
\includegraphics[width=1.65in,height=1.77in]{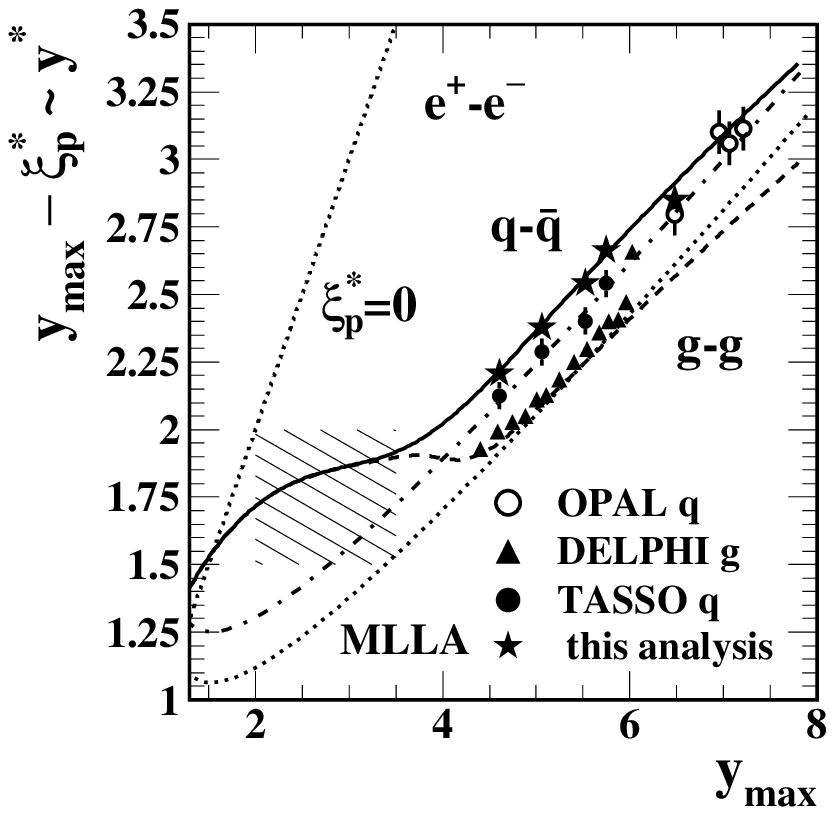}
\includegraphics[width=1.65in,height=1.75in]{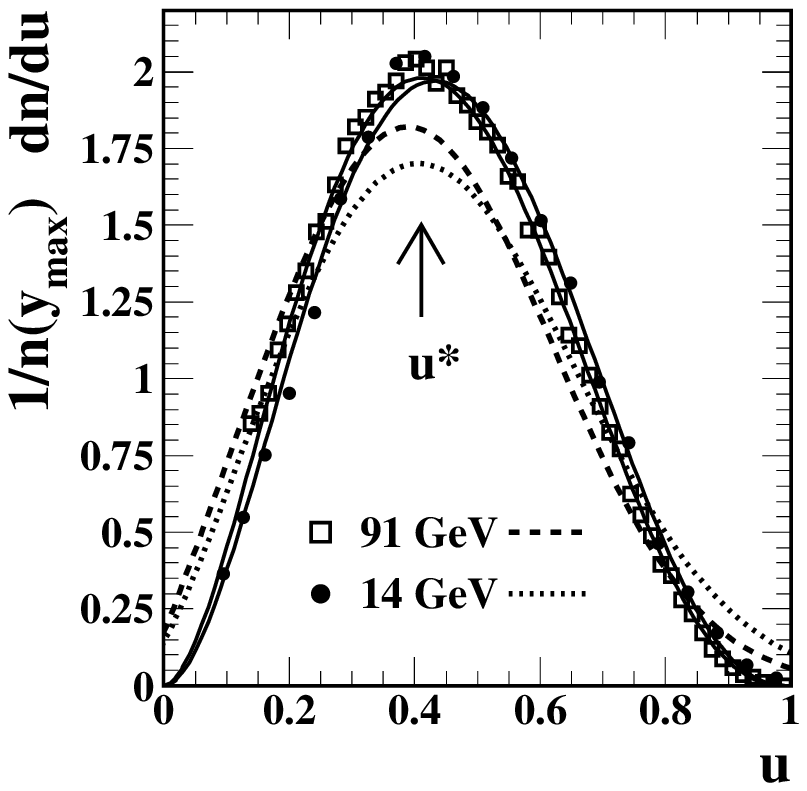}
\caption{\label{fig9}
Left panel: Comparison of fragmentation-function modes {\em vs} parton rapidity from quark and gluon data (points) with `locus of modes' trends (solid and dashed curves) derived from $(p,q)$ energy systematics in Fig.~\ref{fig8} (left panel) and from the MLLA (dash-dot and dotted curves). Right panel: Comparison at two energies of FF data, beta distributions on u and MLLA gaussians suitably transformed to $u$. 
}
\end{figure}


In Fig.~\ref{fig9} (left panel) we show measured values of $\xi_p^*$ in the form $y_{max} -\xi_p^* \sim y^*$ (consistent with Fig.~\ref{fig4}) {\em vs} $y_{max}$ [comparable to plots of $\xi_p^*$ {\em vs} $\ln(Q/\Lambda)$] for eight quark-jet and fourteen gluon-jet energies~\cite{frag,tasso,cacf-delph}. The solid curve $ y^*(y_{max})$  for quark jets inferred from our $(p,q)$ parameterization is the same as the dashed curve in Fig.~\ref{fig14} (left panel). The five stars are obtained from our fits to the fiducial FFs in Table~\ref{betapar1} (compare to peak modes in Fig.~\ref{fig3}). The errors are smaller than the points.

The MLLA predicts for inclusive jets $\xi_p^* = 0.5 Y_2 + c_2 \sqrt{Y_2} - c_2^2$, with $Y_2 \equiv \ln(Q/2\Lambda)$ (note the 2 in the denominator), $c_2 \equiv a/\sqrt{16b\, n_c}$, $a = b / n_c^2$ and $b = (11 n_c - 2 n_f)/3$~\cite{mlla}. The MLLA prediction for $\xi_p^*$ transformed to $y_{max} - \xi_p^*$ is plotted as the dash-dot curve in the left panel. The curve corresponds to $n_f = 5$, but changes with $n_f \rightarrow 3$ are within the data errors. The hatched area is the region of interest for study of low-$Q^2$ partons. The MLLA curve diverges from the $(p_\text{q},q_\text{q})$ parameterization (solid curve) in that region.

We can also obtain a mode prediction for gluon jets. The MLLA prediction for the quark-gluon mode difference is $\Delta \xi^* = \xi_\text{g}^* - \xi_\text{q}^* \approx \frac{1}{12}\left(1+\frac{n_f}{n_c^3}\right) + O(\sqrt{\alpha_s}) \sim 0.1$~\cite{fw-peakstats,cacf-delph}. Taking the inclusive $\xi^*_p$ prediction above as $\xi^*_\text{q}$ we plot $y_{max} - \xi^*_\text{g} = y_{max} - \xi^*_\text{q} - \Delta \xi^*$ as the dotted curve in Fig.~\ref{fig9}, which agrees fairly well above $y_{max} = 4.5$ ($Q \sim 12$ GeV) with the gluon $y^*$ trend (dashed curve) obtained from parameters $(p_\text{g},q_\text{g})$ in Fig.~\ref{fig8}. Data from~\cite{cacf-delph} for FF modes from gluon jets are plotted as solid triangles. The modes were obtained from gaussian fits to gluon FF data over limited intervals on $\xi_p$. The data are well described by the dashed curve obtained from our $(p,q)$ energy systematics and by the MLLA prediction. 



The MLLA width prediction on $\xi_p$ is $\sigma_{\xi_p} = Y^{3/4} / \sqrt{2 c_1}$, with $c_1 = \sqrt{36n_c/b}$~\cite{frag}. The width on $\xi_p$ should be equivalent to the width on $y$ ({\em cf.} Fig.~\ref{fig4}). The variance of the beta distribution on $u$ is $\sigma_u^2 = \frac{p\, q}{(p+q)^2(p+q+1)} \simeq \frac{1}{4(p+q+1)} \sim 0.035$ for flavor-inclusive $e^+$-$e^-$ jets. Thus, the observed r.m.s. width on $y$ is $\sigma_y \sim 0.2\, y_{max} \sim 0.2 Y$, the coefficient {nearly independent} of $y_{max}$ per the $(p,q)$ systematics in Fig.~\ref{fig8}. That result is inconsistent with the MLLA width prediction $\sigma_y \sim 0.37\, Y^{3/4}$. 

Measured FFs have been compared directly with analytical predictions of peak statistics from the MLLA and with gaussians on $\xi_p$ defined by parameters from perturbative approximations~\cite{one-two-particle}. In Fig.~\ref{fig9} (right panel) we compare beta distributions and data for two energies on normalized rapidity $u$ with corresponding MLLA gaussians (normalized to unit integral) using the parameters described above. 
The gaussian tails do not describe the data. Our parameterized model is consistent with pQCD predictions at larger $Q^2$, and the beta distributions (solid curves) demonstrate good sensitivity to small but meaningful systematic variations with energy of the FF data. The good fit of beta distributions to data over all fragment momenta insures a well-defined peak integral. 


\section{Scaling Violations} \label{scalviol}

The na\"ive parton model predicts that parton distribution functions (PDFs) measured in deep inelastic scattering [analogous to FFs $D(x,Q^2)$] should be independent of energy scale $Q^2$ (Feynman-Bjorken scaling)~\cite{naive}. Scaling violations~\cite{heavy}---variations of PDFs and FFs with energy scale---are described by the DGLAP equations~\cite{dglap,dglap2} and depend on the running of strong coupling constant $\alpha_s$, on the available phase space for fragmentation and on $1 \rightarrow 2$ parton splitting described by the Altarelli-Parisi splitting functions. Scaling violations of measured FFs can in turn be used to determine $\alpha_s$ ({\em cf.} Fig.~\ref{fig15}) and to test the predicted values of QCD color factors $C_A$ and $C_F$. In this section we describe scaling violations on conventional momentum/energy fractions and on rapidity variables in terms of the FF parameterization developed in this study.

\subsection{Scaling violations on $(x,Q^2)$}

Fragment distributions on $x_p$ are approximately exponential ({\em cf.} Fig.~\ref{fig2} - left panel). Scaling violations are described in that format as follows. The slope of the FF on $x_p$ becomes more negative (the distribution `softens') with increasing energy scale. Scaling violations (slope change with increasing energy scale) are larger for gluon jets than for quark jets (more radiation is produced by the larger effective color charge of gluons). The strength of scaling violations at large $x_p$ and the accompanying increase of the FF at small $x_p$ (another manifestation of scaling violations) are directly related by energy conservation, as noted in Sec.~\ref{distparam}. Comparisons of quark-jet FFs at different energy scales have been used to measure the running of $\alpha_s$~\cite{webbscalviol,webber-xp-param,alphs-delph}. Comparisons of gluon- and quark-jet FFs have been used to measure the ratio of color factors $C_A / C_F$~\cite{part-pid1,cacf-delph,opal-mult,delphi-scalviol}. 

Scaling violations  are described by QCD theory in the form of the DGLAP equations~\cite{dglap,dglap2}, which are in leading order (LO)
\bea \label{dglapp}
\frac{dD_b(x,s)}{d\ln s} = \frac{\alpha_s(s)}{2\pi} \sum_a \int_x^{1} \frac{dz}{z} \, P_{ab}(z)\, D_a(x/z,s).
\eea
$P_{ab}(z)$ are the Altarelli-Parisi splitting functions~\cite{dglap2}, and $a,b$ denote parton combinations. In a typical study of scaling violations FFs are parameterized at several energy scales $s$ with a model function such as $D(x,s) = N\,x^\alpha\, (1-x)^\beta \, (1 + \gamma/x)$~\cite{webber-xp-param,kkp,cacf-delph}. Such parameterizations can be quite extensive. The KKP parameterization~\cite{kkp} employs 14 parameters for each parton-hadron combination, the energy dependence of each of  $(N,\alpha,\beta,\gamma)$ being described by several polynomial coefficients. The parameters are determined by using the DGLAP equations to evolve the model FFs across energy scales, varying $(N,\alpha,\beta,\gamma)$ with energy to best fit the data and emphasizing the region $x > 0.1$ where pQCD is most applicable ({\em cf.} Fig.~\ref{fig17} - left panel).

\begin{figure}[h]
\includegraphics[width=3.3in,height=1.75in]{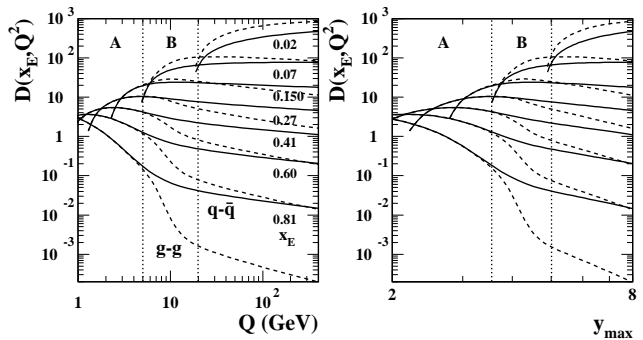}
\caption{\label{fig16}
Scaling violations in a conventional format. Each curve is a conditional slice at fixed $x_E$ from the distribution $2dn/dx_E(x_E,Q^2)$ suitably transformed from model function $2n(y_{max})\, \beta(u\,(y);p,q)$ in Fig.~\ref{fig14} (right panel). 
}
\end{figure}

To illustrate scaling violations in a conventional context with the results of the present study we transform parameterized joint fragment distribution $D(y,y_{max})$ in Sec.~\ref{fragyymax} (Fig.~\ref{fig14}) to $D(x_E,Q^2) = p/(E\, x_E) D[y(x_E, Q),y_{max}(Q)]$. In Fig.~\ref{fig16} (left panel) we plot {conditional} distributions $D(x_E,Q^2)$ for $x_E$ = 0.02, 0.07, 0.15, 0.27, 0.41, 0.60, 0.81 {\em vs} $Q = m_0 \cosh(y_{max})$. The curves for both udsc jets (solid) and gluon jets (dashed) compare well with previous analyses ({\em e.g.,} Fig. 10 of~\cite{part-pid1}). 
The general trends agree with the description of scaling violations noted above but extend over a broader energy range than is usually obtained from data. The sharp falloffs at smaller $Q$ and $x_E$ occur at kinematic limits $\ln x_E \sim y_{min} - y_{max}$ defined by the dotted line in Fig.~\ref{fig14} (right panel). Other features of the distributions correspond to structures in the $(p,q)$ trends of Fig.~\ref{fig8} (left panel). In Fig.~\ref{fig16} (right panel) we replot the same curves {\em vs} $y_{max}$ on a logarithmic scale. Above $y_{max} = 5$ ($Q = 20$ GeV) the curves are nearly straight, revealing the power-law behavior expected for pQCD. 

The vertical dotted lines in both panels separate three regions. Region A ($Q = 1 - 5$ GeV) is dominated by non-perturbative effects. Color charge is effectively hidden (quarks appear similar to gluons) and parton fragment multiplicities are $\sim$ 1 - 3. Although it is least amenable to theoretical treatment, region A produces the majority of parton fragments in nuclear collisions and therefore requires at least a phenomenological characterization consistent with QCD theory. Extrapolation of FF systematics into this region is the purpose of the present analysis. 

Region B ($Q = 5 - 20$ GeV) is the transition region in which color emerges and fragmentation approaches the perturbative description. Convergence of the gluon FF with the quark FF near and below $Q = 10$ GeV in Fig.~\ref{fig16}, determined in this analysis by multiplicity trends alone, is also apparent in direct measurements of scaling violations in FFs~\cite{part-pid1}, for example, deviation of the gluon FF from the HERWIG Monte Carlo at large $x_E$ and for energy scale ($Q_\text{jet} = Q / 2$) 5 - 10 GeV in Fig. 6 of ~\cite{part-pid1}. 

Parameters $(p,q)$ vary weakly and linearly in the energy range above $Q = 20$ GeV ({\em cf.} Fig.~\ref{fig8} -- left panel). If we set the slopes of $p$ and $q$ on $y_{max}$ to zero in that interval the changes in Fig.~\ref{fig16} are small compared to the dominant structure. We conclude that much of the variation in the perturbative region of Fig.~\ref{fig16} is determined by phase-space acceptance variations with parton energy. The subtle linear variations of the $(p,q)$ parameters in that region may provide more differential access to the parton cascade process. To explore that possibility we consider a modified form of the DGLAP equations in the remainder of this section.

\subsection{The running of $\alpha_s(Q)$} \label{runalphas}

\begin{figure}[h]
\includegraphics[width=3.3in,height=1.75in]{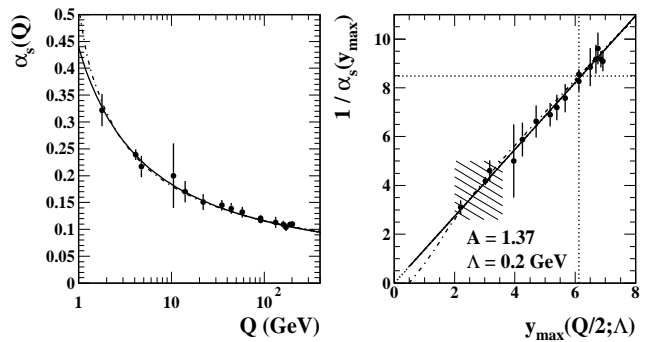}
\caption{\label{fig15}
Left panel: Conventional presentation of $\alpha_s$ systematics on energy scale $Q$. Right panel: Systematics of 1/$\alpha_s$ on rapidity $y_{max}$. The dotted lines cross at $\alpha_s(M_{Z_0})$.  
}
\end{figure}

The energy-dependent $\alpha_s$ factor in the LO DGLAP equations can be approximately eliminated as follows (this procedure reverses use of the energy dependence of scaling violations to infer $\alpha_s$~\cite{alphs-delph}). Data showing the running of $\alpha_s$ with energy scale $Q$ are summarized in Fig.~\ref{fig15} (left panel)~\cite{bethke2,alphas} using a conventional plotting format. A NLO expression for $\alpha_s(Q)$ is~\cite{tung}
\bea \label{alphas}
{\alpha_s}(Q) &=&  \frac{2\pi}{\beta_1 Y} \left\{1 - \frac{\beta_2}{\beta_1^2} \frac{\ln(2Y)}{2 Y}  + O(1)\,\frac{\beta_1}{(2Y)^2} \right\},
\eea
with $Y = \ln(Q/\Lambda)$, $\beta_1 = (11 n_c - 2 n_f)/3$, $n_c$ and $n_f$ the color and effective flavor numbers and $\beta_2 = (102 n_c - 38 n_f)/3$. 
Eq.~(\ref{alphas}) is plotted as the dash-dot curves in Fig.~\ref{fig15}. For $\Lambda = 0.2$ GeV, $n_f = 5$ and $O(1)$ = 0.3 the NLO curves describe the data well. 
The solid curve in the left panel is obtained from the straight-line parameterization in the right panel. 

Fig.~\ref{fig15} (right panel) shows an alternative plotting format in which the data are consistent with a linear trend on rapidity, and the low-$Q^2$ region is visually more accessible. In the leading-log (LL) approximation the strong coupling constant is given by~\cite{dglap2}
\bea
1/\alpha_s(Q) = 1/\alpha_s(Q_0) + \beta_1/2 \pi \cdot \ln(Q/Q_0).
\eea
Since $y_{max}(Q/2;\Lambda) \approx \ln(Q/\Lambda)$ for $Q \gg \Lambda$ we parameterize $\alpha_s$ in the form $ 1/\alpha_s(Q) = A(\Lambda)\, y_{max}(Q/2;\Lambda)$, with the constraint that for any $\Lambda$ the line should pass through the current average $\alpha_s(M_{Z_0}) = 0.118$ (crossed dotted lines). The linear relation is then a single-parameter expression since $A(\Lambda) \approx [1/\alpha_s(M_{Z_0})] / \ln(M_{Z_0}/\Lambda)$. The range of $\Lambda$ values permitted by one-sigma deviations at both the 1.78 and 4.1 GeV data points is $\Lambda = 0.2\pm0.05$ GeV, consistent with the adopted NLO value above. The corresponding $A$ value is $A = 1.37\pm0.05$, which can be compared with the LL slope $\beta_1 / 2\pi = 1.22$ for $n_f = 5$ flavors. The hatched box denotes the region of primary interest for the study of low-$Q^2$ parton fragmentation ($\alpha_s \in [0.2,0.4]$). 
The approximate proportionality of $1/\alpha_s$ and $y_{max}$ apparent in both the LL and NLO descriptions is used in what follows.






\subsection{Scaling violations on ($y,y_{max}$)}


We have determined that $D(x,s) \rightarrow D(\xi,\ln s) \rightarrow D(y,y_{max})$ has a simple underlying structure (beta distribution) and energy dependence [$(p,q)$ parameters]. What are the implications for pQCD and the DGLAP equations? The general behavior of joint distribution $D(y,y_{max})$ is evident from the surface plots in Fig.~\ref{fig14}, where the trivial $\xi$ term in $\ln D(x,s) = \ln D(\xi,s) + \xi$ which dominates Fig.~\ref{fig16} is eliminated, permitting more precise comparisons on a linear scale. Eq.~(\ref{dglapp}) has the form $d D(x,s) /d\ln s \propto \alpha_s(s) \times \text{convolution integral}$. The running coupling constant introduces an energy dependence which can be eliminated. 

As an alternative approach we introduce the {\em logarithmic derivative}~\cite{cacf-delph} motivated by the relation between Mellin transforms of FFs and splitting functions. The Mellin transform of an FF is
\bea
\hat D(w,s) &=& \int_{0}^1 dx\, x^{w - 1}\, D(x,s),
\eea
which is also the Laplace transform of $D(\xi,s)$. The DGLAP equations, written in terms of Mellin transforms, are represented by a simple matrix equation~\cite{dglap2}. For the {\em non-singlet} case $\hat D_\text{ns} = \hat D_\text{q} - \hat D_\text{\=q}$ the logarithmic derivative is
\bea
\frac{d\ln \hat D_\text{ns}(w,s)}{d \ln s} & = & \frac{\alpha_s(s)}{2\pi} \hat P_\text{qq}(w) \equiv \gamma_{qq}(w,s),
\eea
where $\hat P_{qq}(w)$ is the Mellin transform of splitting function $P_{qq}(z)$ for the process q $\rightarrow$ q(z) + g(1-z), with the respective momentum fractions noted, and $\gamma_{ab}(w,s)$ are the {\em anomalous dimensions} of QCD~\cite{gross-qcd,politzer-qcd}.  Since $ d \ln s \approx 2d y_{max}$ we multiply through by $2 y_{max}$ and use the results of the previous subsection to obtain
\bea
\frac{d\ln \hat D_\text{ns}(w,y_{max})}{d \ln y_{max}} & = & \frac{1}{\pi A}\,  \hat P_{qq}(w).
\eea
For the non-singlet case the logarithmic derivative of the Mellin transform of an FF is proportional to the Mellin transform of a splitting function, {\em independent of energy scale} in LO. That simple relation motivates a similar approach to the FFs themselves.

We multiply Eq.~(\ref{dglapp}) by $x \ln (s/m_0^2) / x D(x,s)$ to obtain $d\ln D(\xi,s) / d\ln [\ln (s/m_0^2)] \rightarrow d\ln D(y,y_{max}) / d\ln y_{max}$ on the LHS. The additional factor $\ln (s/m_0^2) \sim 2 y_{max}$ introduced in converting to the logarithmic derivative cancels the $1/y_{max}$ trend of $\alpha_s(y_{max})$ on the RHS to good approximation (Sec.~\ref{runalphas}). What remains on the RHS is the convolution integral, including splitting function $z P(z) = P(\zeta)$ with $\zeta = \ln(1/z)$ and fragmentation-function {ratio} $D_a(\xi - \zeta,s) / D_b(\xi,s)$. We then use the following transformations, $\xi \rightarrow y_{max} - y$, $ \zeta \rightarrow y_{max} - y'$, $D(\xi) \rightarrow  D(y_{max} - \xi) \sim D(y)$ and $D(\xi - \zeta) \rightarrow D(y_{max} - \xi + \zeta) \sim D(y + y_{max} - y')$, to obtain
\bea \label{dglapnew}
\frac{d\ln D_b(y,y_{max})}{d\ln y_{max}} &=& \frac{1}{\pi A} \sum_a \int_y^{y_{max}} \hspace{-.25in} dy' \, P_{ab}(y_{max} - y') \\ \nonumber 
& \times& \frac{D_a(y + y_{max} - y',y_{max})}{D_b(y,y_{max})},
\eea
a modified form of the DGLAP equations on  $(y,y_{max})$.

\begin{figure}[h]
\includegraphics[width=3.3in,height=1.75in]{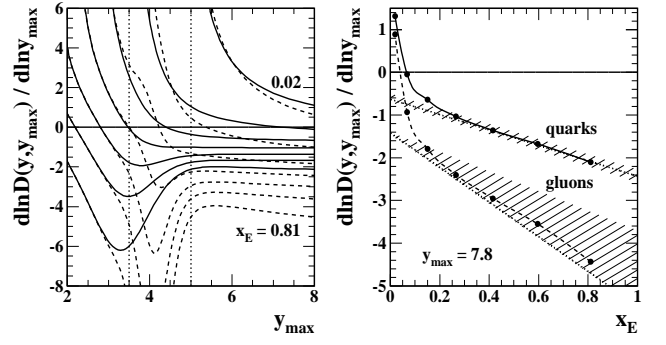}
\caption{\label{fig17}
Left panel: Scaling violations in the form of logarithmic derivatives of the distributions in Fig.~\ref{fig16} (left panel). Solid curves represent udsc quark jets, dashed curves represent gluon jets. The near uniformity to the right of the dotted line ($Q =$ 20 GeV) for larger $x_E$ is notable. Right panel: Logarithmic derivatives for quark and gluon jets at $y_{max} = 7.8$ {\em vs} energy fraction $x_E = \cosh(y) / \cosh(y_{max})$. 
}
\end{figure}

In Fig.~\ref{fig17} (left panel) we plot $d\ln D(y,y_{max})/d\ln y_{max}$ {\em vs} $y_{max}$ for quark (solid) and gluon (dashed) FFs using the parameterized $D(y,y_{max})$ from the present analysis for each parton type. The structure is completely defined by parameters $(p,q)$ in Fig.~\ref{fig8} (left panel) and corresponds exactly to the scaling violations in Fig.~\ref{fig16}. Because of the form of the beta distribution the logarithmic derivative directly relates exponents $p$ and $q$ to the splitting functions. There are three main features in the distributions: 1) nearly-horizontal linear trends at larger energy scales (to the right of the dotted line), 2) `singularities' at smaller energy scales due to kinematic boundaries and 3) minima at intermediate energies corresponding to a transition from `small' (1-2) to `large' (3 or more) fragment number, which may also relate to the emergence of color charge (quark-gluon distinction) at $Q \sim 8$ GeV. The nearly constant values for larger $x_E$ ($y \rightarrow y_{max}$) and $y_{max}$ are the slopes or `power-law' exponents in Fig.~\ref{fig16} (right panel).

In Fig.~\ref{fig17} (right panel) $d\ln D(y,y_{max})/d\ln y_{max}$ {\em vs} $x_E = \cosh(y)/\cosh(y_{max})$ is plotted for quarks and gluons. In the limit $x_E \rightarrow 1$ ($y \rightarrow y_{max}$) and large $y_{max}$ we expect~\cite{cacf-delph} 
\bea
\frac{\{d\ln D(y,y_{max})/d\ln y_{max}\}_\text{gluon}}{\{d\ln D(y,y_{max})/d\ln y_{max}\}_\text{quark}} \rightarrow \frac{C_A}{C_F} = 2.25.
\eea
The dotted lines are $a(x_E + b)$ and $2.25\times a(x_E + b)$, with $a = -1.8$ and $b = 0.35$ adjusted to best match the quark points. The ratio trend is in reasonable agreement with the QCD expectation above $x_E = 0.2$. Within the constraints established in this analysis there is freedom to adjust the energy dependence of $(p_\text{g},q_\text{g})$ to achieve that agreement. Changing the slopes of $(p_\text{g},q_\text{g})$ on $y_{max}$ changes the position of the gluon curve in Fig.~\ref{fig17} (right panel). Variations of $\sim$ 20\% sketched by the lower hatched area are possible without disturbing agreement with data. 

The opposite-sign slopes for $(p_\text{q},q_\text{q})$ in Eq.~(\ref{ppp}) are strongly constrained by the quark-jet multiplicity trend $2n_q(y_{max})$ in Fig.~\ref{fig7}. Setting those slopes to zero gives the lower dotted (quadratic) curve in  the right panel of that figure expected for self-similar FF scaling and fixed mode $u^*$. Varying the same-sign slopes in the $(p_\text{g},q_\text{g})$ expressions together does not affect the gluon-jet multiplicity trend $2n_g(y_{max})$ determined by the $q - p$ difference but does affect the slopes of the gluon logarithmic derivatives at larger $x_E$ in Fig.~\ref{fig17} (left panel) and the slope of the gluon curve in the right panel. Setting the same-sign $(p_\text{g},q_\text{g})$ slopes to zero makes the logarithmic derivative slopes zero, and the gluon curve in the right panel becomes parallel to the quark curve (upper edge of lower hatched region). Increasing the $(p_\text{g},q_\text{g})$ same-sign slopes, which reduces the width of the gluon FF, and hence its amplitude at larger $x_E$, without changing its mode or the multiplicity trend provides a match to the expected $C_A/C_F$ ratio.







\section{Discussion}


Our intention in this study has been to provide the phenomenological means to extrapolate $e^+$-$e^-$ parton fragmentation functions to low $Q^2$ where the perturbative description of QCD is not applicable. We have been guided by the many precise measurements of $e^+$-$e^-$ fragmentation functions and their relationship to pQCD predictions now available. Parton scattering and fragmentation at low $Q^2$ are in turn important for understanding p-p and A-A collisions at RHIC. As a result of this study we have found that the beta distribution provides a simple but precise description of FFs which accomplishes the desired extrapolation but also reveals some interesting new aspects of parton fragmentation.

\subsection{Motivation from p-p and A-A collisions}

We observe large-amplitude two-particle correlations in nuclear collisions at RHIC, produced in part by fragmentation of low-$Q^2$ partons (minijets). Minijet-related correlations observed in p-p collisions are strongly modified with increasing centrality in Au-Au collisions. While low-$Q^2$ partons play an important role in forming the colored medium and driving large-scale hydrodynamic phenomena according to theory, they may also function as sensitive probes of that medium. However, theoretical descriptions of low-$Q^2$ scattering and fragmentation are limited. Factorization is not applicable since low-$Q^2$ parton scattering and fragmentation remain intimately connected. However, new aspects of fragmentation observed {\em via} correlations in nuclear collisions (including strong dependence of fragment angular correlations on $Q^2$~\cite{ppmeas}) suggest a complex but understandable low-$Q^2$ process. 


To facilitate theoretical descriptions of low-$Q^2$ phenomena we have attempted to extrapolate a phenomenological representation of measured FFs in $e^+$-$e^-$ collisions to low $Q^2$. We want to connect two-particle fragment correlations in p-p and heavy ion collisions to pQCD and conventional jet phenomenology through single-particle fragmentation functions. The extrapolation imposes special demands on the fragment representation (particularly for small particle momenta) which have led us to employ rapidity $y$ and normalized rapidity $u$ as our basic kinematic variables. That decision led to the discovery that the beta distribution on $u$ is a good model of light-quark and gluon fragmentation functions.

In p-p and A-A collisions at RHIC we encounter copious parton fragmentation in an energy regime where pQCD is not applicable, but where trends from pQCD may provide semi-quantitative guidance for analysis and interpretation. We therefore distinguish between conditional pQCD fragmentation functions and unconditional fragment distributions measured in nuclear collisions. Given that distinction we can attempt to connect low-$Q^2$ phenomena in nuclear collisions to QCD through the close connection between FFs and fragment distributions as a limiting case. This is a new aspect of fragmentation which lies outside the scope of conventional pQCD fragmentation analysis.

\subsection{The beta distribution as compact representation}

FFs for $e^+$-$e^-$ collisions transformed to $y$ are approximately self-similar with increasing $y_{max}$, as sketched in Fig.~\ref{fig1}. They are bounded by parton rapidity $y_{max}$ and lower limit $y_{min}$. Those trends suggest a further transformation to normalized rapidity $u = (y - y_{min})/(y_{max} - y_{min})$. Measured FFs plotted on $u$ are nearly independent of parton energy and can be factorized into dijet multiplicity $2n(y_{max})$ and unit-normal form factor $g(u,y_{max})$ which is modeled by the beta distribution. Since energy conservation relates the dijet multiplicity to the form-factor shape the fragmentation process is completely represented by the energy-scale dependence of parameters $(p,q)$ of the beta distribution. 

The beta distribution with two energy-dependent parameters thus precisely describes light-flavor FF data over the scale interval $Q \in [5,100]$ GeV ({\em e.g.,} Figs.~\ref{fig2} - \ref{fig5}) and can be extended to lower energies, accomplishing the main goal of this study: extrapolation of fragmentation systematics down to fragment multiplicity $n \sim 1 $-$ 2$ and energy scale $Q \sim 1$ GeV. We also obtain simple representations of FFs on $(x,s)$ or $(y,y_{max})$ (continuous 2D surface) for studies of scaling violations over the full kinematic range. 

\subsection{Fragmentation as an equilibration process}

Why does the beta distribution provide a good description of $e^+$-$e^-$ FFs? $\beta(u;p,q)$ describes systems in which entropy is maximized ({\em e.g.,} by a parton cascade) on a bounded interval ({\em e.g.,} bounded by the leading-parton energy). $\beta(u;p,q)$ is one instance of the {\em exponential family} of probability distributions $p(x)$ which can be defined in terms of a maximum-entropy condition with constraints~\cite{maxent}. The beta distribution defined on $x \in [0,1]$ maximizes the Shannon entropy $S = - \int dx\, p(x) \ln[p(x)]$ subject to constraints on geometric means $ \overline{\ln(x)} = \int dx\, p(x)\, \ln(x)$ and $\overline{\ln(1-x)}$. The gaussian distribution can be similarly defined, with constraints on its first and second moments $\bar x$ and $\overline{x^2}$. 

Given those properties of the beta distribution and its good description of FFs, fragmentation of light quarks and gluons can be viewed as an equilibration process controlled by two opposing tendencies: parton splitting as a form of downscale energy transport which increases entropy and gluon coherence which constrains the splitting at a scale conjugate to hadron size. The observed fragment distribution is then a maximum-entropy configuration balancing those two tendencies. Of the two beta parameters $q$ reflects the splitting tendency and $p$ reflects the hadron size constraint. The DGLAP equations which describe the perturbative splitting (transport) process are thus coupled to the soft part of the FF by entropy maximization as well as energy conservation.  





\subsection{Conventional pQCD fragmentation functions} \label{kkpsec}

\begin{figure}[h]
\includegraphics[width=3.3in,height=1.75in]{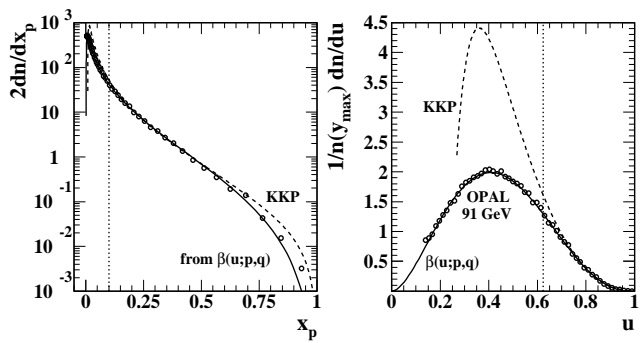}
\caption{\label{fig18}
Left panel: Beta distribution (solid) and KKP FF (dashed) curves compared to OPAL 91 GeV data points (open circles) on linear momentum variable $x_p$. Right panel: The same curves and data transformed to normalized rapidity $u$. The vertical dotted lines both correspond to $x_p = 0.1$. 
}
\end{figure}

Do benefits from the introduction of $(y,y_{max})$, $u$ and the beta distribution justify a new approach to fragmentation? In Fig.~\ref{fig18} we compare FF data to a beta distribution fit from this analysis and a pQCD model FF (KKP) obtained from a conventional scaling violations analysis using the DGLAP equations~\cite{kkp} and defined by 14 parameters for each parton-hadron combination. Fig.~\ref{fig18} (left panel) shows the OPAL 91 GeV FF data from Fig.~\ref{fig2} with the KKP fragmentation function (dashed curve) and the FF from this analysis (solid curve). With the exception of a small deviation at large $x_p$ the agreement in this format appears to be good (also {\em cf.} Fig.~\ref{fig2} -- left panel). 

Fig.~\ref{fig18} (right panel) shows the same distributions on normalized rapidity $u$. The KKP FF based on DGLAP evolution deviates strongly from data below $u = 0.7$ and accurately represents less than 10\% of the fragments at 91 GeV. The vertical dotted line in each panel shows the intended region of validity ($x_p > 0.1$) of the KKP and similar FFs. In contrast, the FF from our analysis, based on the beta distribution and defined by two parameters per parton-hadron combination with simple energy dependence above $Q = 20$ GeV, accurately describes the data on $x_p$ over six orders of magnitude and, extrapolating the full data distribution down to zero momentum, provides a well-defined multiplicity integral. 

Recent improvements in the pQCD description of FFs have lead to much improved coverage on $x_p$. In~\cite{albino} a consistent combination of DGLAP evolution, resummation of soft gluon logarithms and incorporation of hadron mass effects provides a semi-quantitative description below the FF mode on $y$ or $u$ while retaining good agreement above the mode. Such theoretical advances provide a context for the present phenomenological analysis.


\subsection{Energy dependence and scaling violations}


We have studied the energy dependence of FFs as represented by beta distribution parameters $(p,q)$ in Figs.~\ref{fig7} -- \ref{fig9}. The direct relation between dijet multiplicities and FF shape parameters {\em via} the energy sum rule provides a new method for extending the FF description over a broad energy range. 
The connection between shape parameters $(p,q)$ and dijet multiplicity $2n(y_{max})$ is particularly important for extrapolation to low $Q^2$ ({\em e.g.,} the CLEO multiplicity data provide strong constraints on low-energy FF evolution).

A striking feature of Fig.~\ref{fig7} is the bifurcation of quark and gluon trends near $Q \sim 5$ GeV and evolution to large separation by $Q \sim 20$ GeV, with a dramatic correspondence in Fig.~\ref{fig8} (left panel). Above 20 GeV the system exhibits simple pQCD trends which are most apparent in Fig.~\ref{fig8}. Corresponding peak modes on $u$ remain close to 0.4 over a broad energy range. It is not clear from Fig.~\ref{fig7} that peak modes increase rapidly near 5 GeV and approach unity below that point, but that trend can be inferred from Fig.~\ref{fig8} which provides important extrapolation guidance. Mode trends are illustrated in Fig.~\ref{fig9} and agree well with data and pQCD predictions in the energy range above 15 GeV. 

We have considered scaling violations in Fig.~\ref{fig14} and Figs.~\ref{fig16} - \ref{fig17}. We find that a combination of QCD scaling violations and gluon coherence leads to approximate {\em self-similarity} of FFs on rapidity with changing energy scale, as illustrated in Fig.~\ref{fig1} and demonstrated in Fig.~\ref{fig3}. Normalized FFs plotted on normalized rapidity $u$ are therefore {nearly} independent of energy scale. However, relative to the self-similarity trend we observe more subtle forms of `scaling violations' as described in the previous paragraph. 

Conventional scaling-violation systematics are easily and precisely reproduced by our parameterization over a broad energy range, as demonstrated in Fig.~\ref{fig16} down to kinematic limits. It is straightforward to explore the consequences of varying $(p,q)$ energy trends. For example, Fig.~\ref{fig17} (right panel) demonstrates that the $C_A/C_F$ limit for logarithmic derivatives previously established by specific experimental measurements ({\em e.g.,} ~\cite{part-pid1,cacf-delph,opal-mult,delphi-scalviol}) is consistent with the $(p,q)$ parameterization determined by the present study. 

\section{Summary}

In conclusion, we have examined the variation of fragmentation functions (FF) with parton energy on conventional kinematic variables $x_p$ and $\xi_p$, on rapidity $y$ and on normalized rapidity $u$. We find that FFs plotted on rapidity $y$ vary with energy in a nearly self-similar manner. FFs transformed to $u$ are well described by a product of the dijet multiplicity and a unit-normal form factor modeled by the beta distribution. The latter is determined by parameters $(p,q)$ which exhibit modest linear variations within perturbative energy scale range $Q > 20$ GeV. The beta distribution shape, when combined with an energy-conservation sum rule, also determines FF multiplicities. The factored representation on $u$ thus provides a simple and compact representation of $e^+$-$e^-$ FFs over a broad energy range and permits extrapolation to small energy scales.

Reduction of scaling violations to near-linear variations of beta parameters $(p,q)$ may provide more differential access to the energy dependence of parton fragmentation. The beta distribution model suggests that fragmentation of light partons can be viewed at larger energy scales as an entropy-maximizing equilibration process. In this analysis the energy dependence of fragmentation has been extrapolated down to a kinematic region not accessed by conventional methods. Such low-$Q^2$ extrapolation provides a phenomenological context for minijet-related two-particle correlations in p-p and A-A collisions at RHIC, forming a basis for theoretical treatments of in-medium dissipation of low-$Q^2$ partons and the subsequent hadronization process in heavy ion collisions.

This work was supported in part by the Office of Science of the U.S. DoE under grant DE-FG03-97ER41020.


\end{document}